\documentclass[prb,superscriptaddress,twocolumn]{revtex4-2} \usepackage{silence}
\usepackage{times}
\usepackage{color}
\usepackage[colorlinks=true,urlcolor=blue,citecolor=blue]{hyperref}
\usepackage{url}
\usepackage{breakurl}
\usepackage{soul}
\usepackage{graphicx}
\usepackage{amsmath}
\usepackage{subfigure}
\usepackage{xcolor}
\usepackage{bibunits}
\usepackage{amssymb}
\usepackage{latexsym}
\DeclareGraphicsExtensions{.png,.eps}
\usepackage{float}
\usepackage{appendix}
\usepackage{etoolbox}
\usepackage{physics}

\usepackage{xcolor}
\hypersetup{
    colorlinks = true,                    
    citecolor = {blue},
    linkcolor = {purple},
}

\begin{document}

\title{Statistics-encoded tensor network approach in disordered quantum many-body spin chains}
\author{Hao Zhu}
\affiliation{School of Physics, Beihang University,100191,Beijing, China}
\author{Ding-Zu Wang}
\email[Corresponding author: ]{dingzu\_wang@sutd.edu.sg}
\affiliation{Science, Mathematics and Technology Cluster, Singapore University of Technology and Design, 8 Somapah Road, 487372 Singapore}
\affiliation{Centre for Quantum Technologies, National University of Singapore 117543, Singapore}
\author{Shi-Ju Ran}
\email[Corresponding author: ]{sjran@cnu.edu.cn}
\affiliation{Center for Quantum Physics and Intelligent Sciences, Department of Physics, Capital Normal University, Beijing 10048, China}
\author{Guo-Feng Zhang}
\email[Corresponding author: ]{gf1978zhang@buaa.edu.cn}
\affiliation{School of Physics, Beihang University,100191,Beijing, China}

\begin{abstract}
Simulating the dynamics of quantum many-body systems with disorder is a fundamental challenge.
In this work, we propose a general approach---the statistics-encoded tensor network (SeTN)---to study such systems.
By encoding disorder into an auxiliary layer and averaging separately, SeTN restores translational invariance, enabling a well-defined transfer matrix formulation. 
We derive a universal criterion, $n \gg \alpha^2 t^2$, linking discretization $n$, disorder strength $\alpha$, and evolution duration $t$. 
This sets the resolution required for faithful disorder averaging and shows that encoding is most efficient in the weak-disorder, typically chaotic regime.
Applied to the disordered transverse-field Ising model, SeTN shows that, over the numerically accessible time window, the spectral form factor is governed by the leading transfer-matrix eigenvalue, in contrast to the kicked Ising model.
SeTN thus provides a novel framework for probing the disorder-driven dynamical phenomena in many-body quantum systems. 
\end{abstract}

\maketitle

\section{Introduction}

Simulating quantum many-body systems with disorder is crucial for understanding a wide range of phenomena, including quantum chaos~\cite{gutzwillerChaosClassicalQuantum1990a,dalessioQuantumChaosEigenstate2016b,haakeQuantumSignaturesChaos2018}, many-body localization (MBL)~\cite{nandkishoreManyBodyLocalizationThermalization2015c,abaninColloquiumManybodyLocalization2019b,sierantManybodyLocalizationAge2025}, quantum spin glass~\cite{brayReplicaTheoryQuantum1980,sachdevGaplessSpinfluidGround1993}, and criticality~\cite{fisherCriticalBehaviorRandom1995,pichCriticalBehaviorGriffithsMcCoy1998}.  
Recently, the chaotic dynamics of disordered systems predicted by random matrix theory (RMT) has attracted growing interest, driven by advances in experimental techniques~\cite{liEmergentUniversalQuench2024,dongMeasuringSpectralForm2025}.

Analytical progress on disorder-averaged dynamics has primarily been achieved in random quantum circuits and dual-unitary models, where Haar averaging or self-duality enables exact transfer-matrix treatments~\cite{chanSolutionMinimalModel2018c,bertiniExactSpectralForm2018,vonkeyserlingkOperatorHydrodynamicsOTOCs2018a,zhouEmergentStatisticalMechanics2019,nahumOperatorSpreadingRandom2018c,sonnerCharacterizingManybodyLocalization2022,riddellStructuralStabilityHypothesis2024,bertiniExactlySolvableManybody2025a}. 
However, these approaches cannot be directly extended to generic time-independent Hamiltonians, where self-duality is absent and the transfer-matrix dimension grows exponentially, hindering practical implementation.
Numerically, exact diagonalization (ED) is restricted to small system sizes~\cite{pietracaprinaShiftinvertDiagonalizationLarge2018a,sierantPolynomiallyFilteredExact2020a}, while conventional tensor-network methods and strong-disorder renormalization approaches~\cite{doggenManybodyLocalizationDelocalization2018a,chandaTimeDynamicsMatrix2020,hikiharaNumericalRenormalizationgroupStudy1999,goldsboroughSelfassemblingTensorNetworks2014a} 
treat each disorder realization independently and are applicable mostly to the deep MBL regime.
Another widely used route in this regime, quantum-parallel method~\cite{alvarezQuantumParallelismTool2008a,paredesExploitingQuantumParallelism2005c,andraschkoPurificationManyBodyLocalization2014,hubigTimedependentStudyDisordered2019a}, embeds disorder realizations taking only discrete values into an auxiliary system and evolves them in parallel. 
However, when the disorder follows a continuous distribution, as is typical in studies of chaotic dynamics, the method requires an auxiliary space of infinite dimension.

In this letter, we propose the statistics-encoded tensor network (SeTN), a general approach for studying disordered many-body systems with time-independent Hamiltonians.
In SeTN, we encode disorder into an auxiliary layer on which the average is taken independently at each site (Fig.~\ref{fig:tn_repr}(b)).  
By analyzing the decay rate of the singular values, we derive a universal criterion for the temporal resolution required for efficient disorder encoding: $n \gg \alpha^2 t^2$, where $n$ is the number of time steps, $\alpha$ the disorder strength, and $t$ the evolution duration.
This criterion shows that continuous disorder distributions can be efficiently encoded, particularly in the weak-disorder regime, making disorder-averaged transfer matrices numerically feasible for generic Hamiltonians.

After averaging, SeTN restores spatial translation invariance, yielding a compact MPO transfer matrix that can be directly analyzed using Krylov iteration~\cite{stewartKrylovSchurAlgorithmLarge2002}  and density matrix renormalization group (DMRG) methods~\cite{whiteDensityMatrixFormulation1992a,whiteDensitymatrixAlgorithmsQuantum1993}.
As a demonstration, we investigate the spectral form factor (SFF) of the disordered transverse-field Ising model in its chaotic regime, revealing, within the numerically accessible regime, a prolonged pre-RMT transient dominated by a single, non-degenerate leading eigenvalue of the transfer matrix.
This pathway to RMT behavior contrasts with Floquet models such as the kicked Ising model, where chaotic spectral statistics and eigenvalue degeneracies appear from the outset~\cite{chanSolutionMinimalModel2018c}.

\section{Statistics-encoded tensor network approach}

\begin{figure}[tpb]
    \centering
    \includegraphics[width=\linewidth]{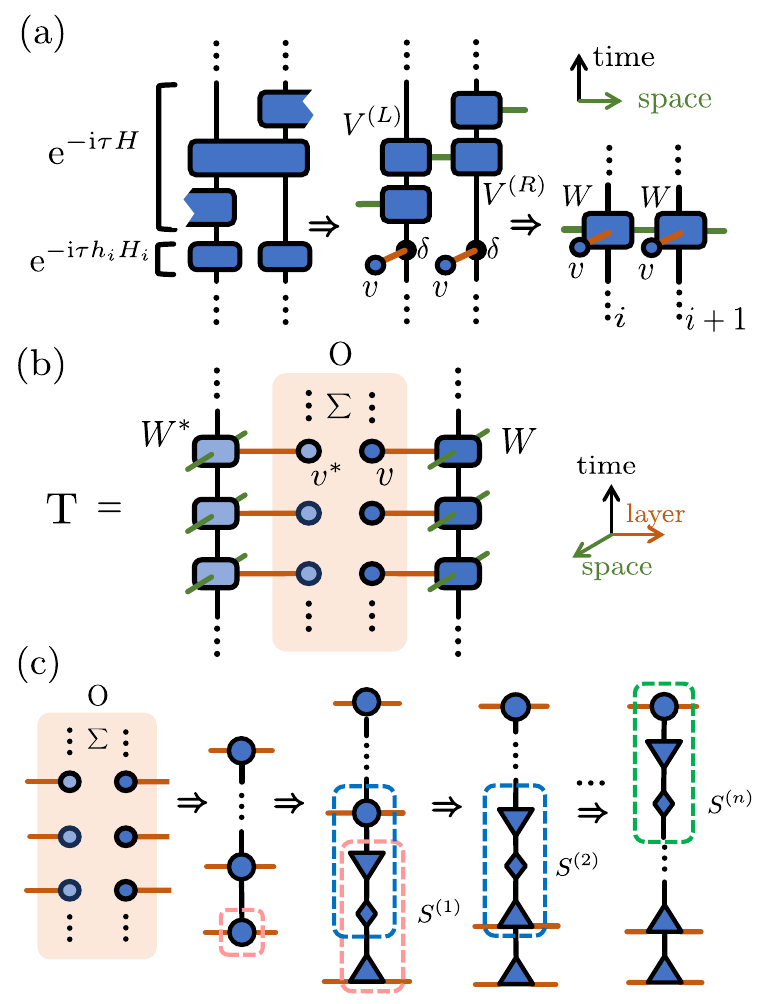}
    \caption{(Color online)
    (a) Tensor network (TN) representation of the Trotterized evolvation operator. The tensor encoding the disorder is decomposed into a vector (blue circle) and a Kronecker delta tensor (black point). 
    (b) A general two-layer TN representation of the disorder-averaged transfer matrix $\mathrm{T}$. The orange-shaded region indicates the disorder average.
    (c) Diagrammatic illustration of the SeTN compression. $ S^{(n)} $ is the singular values obtained via corresponding singular value decomposition.
    }
    \label{fig:tn_repr}
\end{figure}

We consider a quantum many-body system with local disorder, described by the Hamiltonian $\mathcal{H} [\boldsymbol{h}] = H + \sum_{i = 1}^{L} h_i H_{i}$,
where $H$ is the disorder-free part, $\{H_{i}\}$ are local operators acting on site $ i $, $\{h_i\}$ are disorder variables which are independent and identically distributed, and $L$ is the system size.
Throughout this work we set $\hbar = 1$.
To simulate the real-time dynamics, we apply a first-order Trotter decomposition with a small time step $\tau \ll 1$, which gives the evolution operator 
$U(\tau) = \mathrm{e}^{ -\mathrm{i} \tau H } \mathrm{e}^{ - \mathrm{i} \tau \sum_i h_{i} H_{i} } + O(\tau^2)$.
This decomposition enables a TN representation of the full evolution operator (see Fig.~\ref{fig:tn_repr}(a)). 
For simplicity, we focus on the widely studied case where $H = \sum_i H_{i,i + 1}$ includes only nearest-neighbor interactions and $H_{i} = \sigma^{z}_{i}$ is the Pauli-$z$ operator on site $i$.
Generalizations to longer-range interactions or different disorder couplings are straightforward~\cite{supplementaryMaterial}.

To proceed, as shown in Fig.~\ref{fig:tn_repr}(a), we first decompose the two-body unitary gates into two local tensors,
\begin{equation} 
    \begin{aligned}
        \langle j_n^{[i]}j_{n+1}^{[i]} \vert \mathrm{e}^{-\mathrm{i}\tau H_{i,i + 1}} \vert &j_n^{[i + 1]}j_{n+1}^{[i + 1]} \rangle
        = V^{(L)}_{j_n^{[i]}j_{n+1}^{[i]},k_n^{[i]}} V^{(R)}_{k_n^{[i]},j_n^{[i + 1]} j_{n+1}^{[i + 1]}},
    \end{aligned}
\end{equation}
where $j_n^{[i]}$ refers to site $i$ and time step $n$, with repeated indices summed over implicitly.
This can be achieved via singular value decomposition (SVD) or spectral decomposition.
At the same time, we factor out the diagonal disorder term using a delta tensor: $\langle j_n^{[i]} \vert \mathrm{e}^{-\mathrm{i}\tau \sigma^{z}_{i}} \vert j_{n+1}^{[i]}\rangle = v_{l_n^{[i]}}\delta_{l_n^{[i]} j_n^{[i]} j_{n+1}^{[i]}}$,
where $\delta_{ijk}$ is the three-order Kronecker tensor (equal to $1$ when $i=j=k$ and $0$ otherwise), 
and $ v_{l} = \mathrm{e}^{-\mathrm{i}\tau h_i l} $ with $ l \in \{1, - 1\} $ encodes the effect of local disorder.
Throughout this work, we use the indices $j,k$ and $l$ to denote the \textit{time}, \textit{space}, and \textit{layer} directions, respectively, as indicated by the coordinate system in Fig.~\ref{fig:tn_repr}.
By combining these decompositions, we define a local tensor $W$ as
\begin{equation} 
    W_{j_n^{[i]}j_{n+1}^{[i]},k_n^{[i - 1]}k_n^{[i]},l_n^{[i]}} = 
    \delta_{l_n^{[i]} j_n^{[i]} p_n^{[i]}}
    V^{(R)}_{k_n^{[i - 1]},p_n^{[i]} q_{n}^{[i]}}
    V^{(L)}_{q_n^{[i]}j_{n+1}^{[i]},k_n^{[i]}}, 
\end{equation}
where $p$ and $q$ are auxiliary indices used for internal contractions.
This construction is illustrated in the third column of Fig.~\ref{fig:tn_repr}(a), where the spatial slice at site $i$ corresponds to $\prod_{n} v_{l_{n}^{[i]}} W_{j_n^{[i]}j_{n+1}^{[i]},k_n^{[i - 1]}k_n^{[i]},l_n^{[i]}}$.

\begin{figure*}[tpb]
    \centering
    \includegraphics[width=\linewidth]{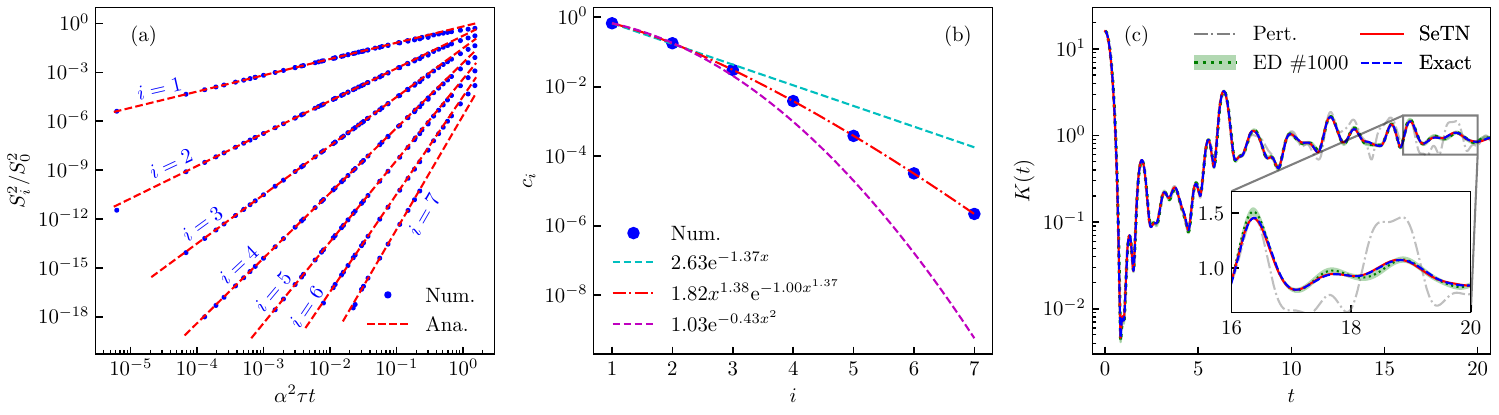}
    \caption{(Color online)
    (a) Blue dots: squared relative singular values from SeTN compression under various parameters (Num.). Dashed lines: analytic predictions from perturbation theory (Ana.), with coefficients fitted to numerical data.
    (b) Blue dots: first seven coefficients $c_i$ obtained by fitting the data in (a). These are further fitted to exponential (green dashed), Gaussian (purple dashed), and gamma-like (red dot-dashed) functions.
    (c) Spectral form factor of the disordered TFIM with system size $L = 4$, computed using four methods: exact diagonalizing 1000 disorder realizations (avg \#1000, green dotted; shaded band indicates the standard error), numerical integration (Exact, blue dashed), SeTN (red solid), and perturbative results (Pert., gray dot-dashed). The mean deviation between SeTN and the exact result is approximately 0.021. Parameters: $J = b = 1$, Trotter step $\tau = 0.005$, averaging over $M = 10^6$ realizations; singular values below $10^{-10}$ are truncated.
    }
    \label{fig:sv_scale}
\end{figure*}

We study disorder-averaged dynamics, for which many quantities of interest including averaged operator expectation values $\langle\!\langle O(t) \rangle\!\rangle = \langle\,\operatorname{Tr}[O\,U(t)\rho(0)U^\dagger(t)]\,\rangle$, as well as the SFF $K(t) = \langle\,\operatorname{Tr}[U(t)\otimes U^\dagger(t)]\,\rangle$, naturally admit a two-layer tensor-network representation obtained by contracting the forward and backward evolutions $U$ and $U^\dagger$ (dark/light blue in Fig.~\ref{fig:tn_repr}(b)).
The effect of disorder is encoded in a statistics layer (orange region) which, after averaging, renders all spatial sites equivalent. This structure motivates the definition of a transfer matrix $\mathrm{T}$ that captures the combined action of unitary dynamics and disorder averaging across a single spatial slice,
\begin{equation}
\begin{aligned}
    &\mathrm{T}_{k_1^{\prime[i - 1]},...,k_n^{\prime[i - 1]},k_1^{[i - 1]},...,k_n^{[i - 1]};k_1^{\prime[i]},...,k_n^{\prime[i]},k_1^{[i]},...,k_n^{[i]}} \\
    &= 
    \Big\langle
    \prod_{p = 1}^{n} W^*_{j_p^{\prime[i]}j_{p+1}^{\prime[i]},k_p^{\prime[i - 1]}k_p^{\prime[i]},l_p^{\prime[i]}} v^*_{l_{p}^{\prime[i]}} v_{l_{p}^{[i]}} W_{j_p^{[i]}j_{p+1}^{[i]},k_p^{[i - 1]}k_p^{[i]},l_p^{[i]}}
    \Big\rangle
    \\ &= \mathrm{O}_{l'_1,...,l'_n;l_1,...,l_n} \prod W^* W,
\end{aligned}
\end{equation}
where the indices on the conjugated tensors are distinguished by primes and $\langle{\cdot}\rangle$ denotes the disorder average.

In the last line, We separate the disorder-averaged tensor $\mathrm{O}$ from the unitary layers (omitting site superscripts and tensor subscripts for clarity), which can be obtained explicitly by integrating over the distribution of $h$:
\begin{equation}\label{eq:O}
\begin{aligned}
    &\mathrm{O}_{l'_1,...,l'_n;l_1,...,l_n} \equiv \Big\langle{\prod_{p = 1}^{n}v_{l'_p}^* v_{l_p}}\Big\rangle
    = \int \mathrm{d} h \; \mathrm{e}^{ \mathrm{i} h \tau \sum_{p}^{} (l'_{p} - l_{p}) } P(h)
    \\ 
    &=
    \begin{cases}
        \operatorname{sinc} \left[ \tau \alpha \sum_{p}^{} (l'_{p} - l_{p}) \right], & h \sim \mathcal{U}(-\alpha, \alpha), \\[4pt]
        \exp \left[ 
        - \frac{1}{2} \sigma^2 \tau^2 
        \left( 
        \sum_{p}^{} 
        ( 
            l'_{p} - l_{p}
        )
        \right)^2
    \right], & h \sim \mathcal{N}(0, \sigma^2).
    \end{cases}
\end{aligned}
\end{equation}
where $P(h)$ denotes the normalized probability density function of the disorder variable $h$, i.e., $\int dh\, P(h)=1$.
We give explicit formulas for both uniform $\mathcal{U}(-\alpha,\alpha)$ and Gaussian $\mathcal{N}(0,\sigma^2)$ disorder. 
While a similar expression for the Gaussian case was derived in the KIM context~\cite{bertiniExactSpectralForm2018}, the present formulation applies to general static Hamiltonians with arbitrary on-site disorder.
Moreover, when $P(h)$ is discrete, $\mathrm{O}$ reduces exactly to the standard quantum-parallel encoding~\cite{supplementaryMaterial}.

Although Eq.~\ref{eq:O} represents the ideal analytic average, directly constructing the $2^n\times2^n$ tensor $\mathrm{O}$ is numerically infeasible.
In practice, we therefore adopt a finite-sample approximation with $M$ disorder realizations sampled from the distribution $P(h)$, and compress the resulting MPO through the two-step procedure illustrated in Fig.~\ref{fig:tn_repr}(c).
First, each Se-Layer realization is written as a matrix product operator (MPO) with bond dimension 1; summing over $M$ such realizations via MPO addition~\cite{crosswhiteFiniteAutomataCaching2008} yields an MPO of bond dimension $M$ encoding the disorder average.
Second, we iteratively compress this MPO using SVD.
At $n=1$ (pink box in Fig.~\ref{fig:tn_repr}(c)), 
the singular values $S^{(1)}$ below a fixed threshold are discarded.
The truncated tensors are contracted with $A^{(2)}$ and decomposed again (blue box in Fig.~\ref{fig:tn_repr}(c)), 
and the process continues site by site to the final layer.

Notely, since each step involves only local operations and the tensors $A^{(n)}$ are identically constructed via MPO addition from individual realizations, the overall memory requirement reduces to $O(M)$~\cite{supplementaryMaterial}, compared to the naive $O(n M^2)$ scaling.
This significantly enhances computational efficiency and enables averaging over a large number of realizations ($M \gg 1$) in practice.


\section{Properties of Se-Layer and SeTN effectiveness}

The previous section outlined the implementation of SeTN.
A natural question is whether disorder averaging can be faithfully encoded into the TN representation.
This effectiveness crucially depends on the behavior of the singular value spectrum $S^{(n)}$, illustrated in Fig.~\ref{fig:tn_repr}(b): only when the singular values decay rapidly, ideally exponentially, can the SeTN accurately compress the averaged transfer matrix with a low bond dimension.
To analyze this, we derive an analytic expression for the spectrum by solving the recursive structure in SeTN compression perturbatively~\cite{supplementaryMaterial}:
\begin{align}\label{eq:sv_scale}
    \left(S_i(t)/S_0(t)\right)^2 \simeq c_i \left(\alpha^2 \tau t\right)^i,
\end{align}
where $S_i(t)$ denotes the $i$-th singular value at the $n$-th compression step (with $n=t/\tau$), ordered in descending magnitude, $S_0(t)$ is the largest and $\alpha$ is the disorder strength.
The coefficients $c_i$ depend only on the disorder statistics; for a uniform distribution, the first two values can be computed exactly as $c_1 = 2/3$ and $c_2 = 8/45$~\cite{supplementaryMaterial}.
This result shows that the decay of the singular values is governed by the dimensionless parameter $\alpha^2 \tau t$, which confirms the exponential suppression of subleading components in the singular value spectrum when $\alpha^2 \tau t \ll 1$, or equivalently,
\begin{align}\label{eq:sv_scale}
    n \gg \alpha^2 t^2.
\end{align}
This condition ensures that SeTN achieves accurate, low-bond-dimension compression of disorder-averaged transfer matrices.

To validate this analytic result, we numerically investigate the singular value spectra under various parameter settings~\cite{supplementaryMaterial} and compare them with the perturbative result. 
As shown in Fig.~\ref{fig:sv_scale}(a), the blue dots represent the normalized squared singular values $(S_i(t)/S_0(t))^2$ obtained from numerical simulations using a uniform disorder distribution, while the red dashed lines show the corresponding analytic estimates.
The excellent agreement between the numerical data and theoretical predictions in the region $\alpha^2 \tau t \ll 1$ confirms the exponential decay behavior discussed above.

In Fig.~\ref{fig:sv_scale}(b), we further analyze the coefficients $c_i$ appearing in Eq.~\ref{eq:sv_scale}.
The plot shows the first seven coefficients extracted by fitting the numerical spectra in Fig.~\ref{fig:sv_scale}(a) to the analytic form. 
Notably, these coefficients decay faster than exponentially but slower than Gaussian, and are well approximated by a gamma-like function, which highlights the intrinsic efficiency of the SeTN method.
Importantly, despite the rapid decay, the leading coefficients remain large. 
For a typical setting where $\alpha^2 \tau t = 10^{-2}$ (e.g., $\alpha = 0.1$, $\tau = 0.01$, $t = 100$), the corresponding contribution to $(S_i/S_0)^2$ at $i = 7$ is on the order of $10^{-14}$, several orders of magnitude smaller than the coefficient $c_7 = 10^{-6}$.
Thus, the overall compression efficiency of SeTN is still primarily governed by the dimensionless combination $\alpha^2 \tau t$, as emphasized earlier.

For fixed disorder strength $ \alpha $ and time $ t $, we provide a physical interpretation of the above properties: while increasing the discretization number $ n $ leads to exponential growth in the dimension of the transfer matrix, the total information introduced by disorder averaging, quantified by the dimensionless parameter $\alpha^2 t^2$, remains unchanged.
In the SeTN approach, this information is distributed across a tensor network of length $n$, such that the amount of information encoded per time step scales as $\alpha^2 t^2/n = \alpha^2 \tau t$.
This local information density is directly related to the decay of the singular value spectrum, which allows efficient compression and accurate representation of disorder-averaged dynamics when $n \gg \alpha^2 t^2$.

\begin{figure}[tpb]
    \centering
    \includegraphics[width=\linewidth]{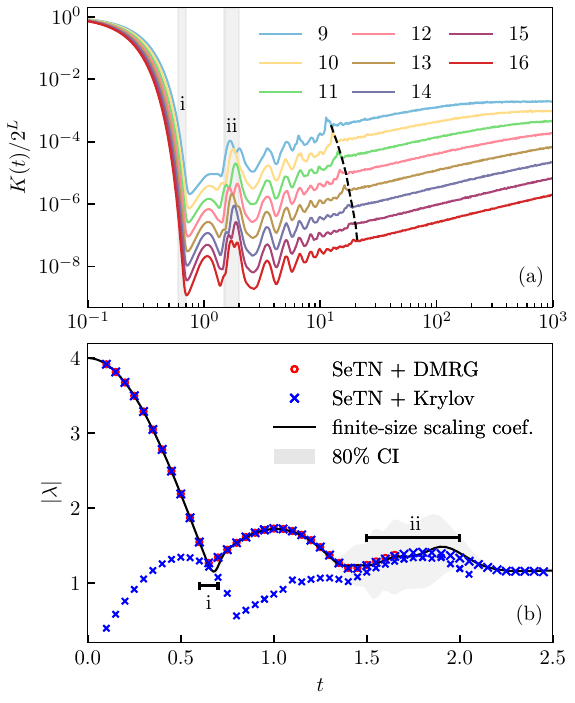}
    \caption{(Color online)
        (a) SFF of the disordered TFIM with sizes $L = 9$ to $16$ (top to bottom), computed by ED using 2440 independent disorder realizations per size. Standard errors are below 4\% at all times, and are therefore not visible on the logarithmic scale. Shaded regions (i) and (ii) indicate the first and second rebounce windows.
        (b) Magnitudes of the leading and subleading eigenvalues of the SeTN-derived transfer matrix as functions of time, obtained using Krylov (blue crosses) and non-Hermitian DMRG (red circles).
        Horizontal bars (i) and (ii) match the time windows in (a), highlighting the near-degeneracy of dominant eigenvalues. 
        Black lines represent the fitted coefficient extracted by fitting the data in panel (a) to the form $\lambda(t)^L$, and the shaded region indicates the 80\% confidence interval. Parameters: Trotter step $\tau = 0.05$, $J = b = 1, \alpha=0.5$.
    }
    \label{fig:largest_eig}
\end{figure}

\section{Scaling behavior of spectral form factor}

The preceding analysis shows that SeTN is more efficient for weak disorder, as singular values decay faster. 
In this regime, the system is typically driven into chaos, which is challenging for conventional tensor networks.
To illustrate SeTN’s ability to handle chaotic dynamics, we investigate the spectral form factor (SFF) of the disordered transverse-field Ising model (disordered TFIM) in its chaotic regime.

The SFF, defined as the Fourier transform of the two-point correlation function of the spectral density~\cite{supplementaryMaterial},
\begin{equation}\label{eq:sff}
K(t) = \langle{ |\operatorname{Tr} U(t)|^2 }\rangle = \operatorname{Tr}
\langle{ U(t) \otimes U^*(t) }\rangle,
\end{equation}
where $U(t) = \mathrm{e}^{-\mathrm{i}H t}$ is the evolution operator, serves as one of the simplest analytically tractable probes of quantum chaos.
It captures both short- and long-range spectral correlations and has been extensively used in many-body chaos studies~\cite{chanSolutionMinimalModel2018c,kosManyBodyQuantumChaos2018b,chanSpectralStatisticsSpatially2018c,bertiniExactSpectralForm2018,dongMeasuringSpectralForm2025}, where it typically exhibits an initial transient followed by the ramp–plateau structure predicted by RMT~\cite{haakeQuantumSignaturesChaos2018,liuSpectralFormFactors2018b}.

Here, we study the SFF in the disordered TFIM whose Hamiltonian reads, $\mathcal{H} [\boldsymbol{h}] = J \sum_{i=1}^{L-1} \sigma^{z}_{i} \sigma^{z}_{i+1} + b \sum_i^L \sigma^x_{i} + \sum_i h_i \sigma^{z}_{i}$, where $J$ is the coupling strength, $b$ is the magnetic field, and $h_i$ are independent random variables drawn from a uniform distribution in the range $[-\alpha, \alpha]$.
We use open boundary conditions throughout; the results are insensitive to the choice of boundary condition.
Setting $J = b = 1$ and $\alpha = 0.5$ where the system is chaotic~\cite{supplementaryMaterial}, we compute the SFF via ED for $L = 9$ to $16$ averaging $2440$ independent disorder realizations for each size.
The results clearly display the onset of the linear ramp as shown in Fig.~\ref{fig:largest_eig}(a).

Since the SFF is non-self-averaging~\cite{prangeSpectralFormFactor1997b}, reliable evaluation typically requires extensive disorder averaging. 
By expressing the SFF into a two-layer tensor-network form, however, SeTN can be directly employed as outlined in the previous section.
Fig.~\ref{fig:sv_scale}(c) shows that SeTN (red line) reproduces converged results obtained from numerical integration (blue dashed line) even at long times.
For comparison, we include disorder-averaged ED ($10^3$ samples, green dotted) and second-order perturbation results (SeTN with bond dimension three~\cite{supplementaryMaterial}, gray dash-dotted). 
The slight ED deviation (see inset of Fig.~\ref{fig:sv_scale}(c)) highlights SeTN’s accuracy, whereas the breakdown of perturbation theory emphasizes the importance of small singular values for long-time dynamics.

A key advantage of SeTN is its ability to restore spatial translational invariance broken by disorder, thereby enabling a well-defined transfer matrix and access to the thermodynamic limit.
In Fig.~\ref{fig:largest_eig}(b), we show the largest eigenvalues of the transfer matrix, computed by Krylov iteration (blue crosses) and non-Hermitian DMRG (red circles)~\cite{supplementaryMaterial}.
For reference, we fit the finite-size ED data in Fig.~\ref{fig:largest_eig}(a) to the form $K(t) = \lambda(t)^L$, with the extracted $\lambda(t)$ plotted as black lines.
The excellent agreement between the largest eigenvalues and $\lambda(t)$ demonstrates that, within the accessible pre-ramp time window, the SFF is indeed governed by the non-degenerate leading eigenvalues of the transfer matrix.
This behavior contrasts with Floquet models such as the kicked Ising model~\cite{chanSolutionMinimalModel2018c}, where the SFF is controlled by a highly degenerate set of leading eigenvalues from the outset.

Within the accessible pre-ramp time window, the dynamics are governed by the non-degenerate leading eigenvalue, which is size independent, cannot explain the size-dependent onset of the RMT ramp (the Thouless time, black dashed in Fig.~\ref{fig:largest_eig}(a)).
Understanding how the system crosses over from the leading-eigenvalue regime to the ramp regime thus requires access to longer times.

At present, a direct evaluation of the eigenvalues in the ramp region is hindered by rapid bond growth in Krylov iteration and the lack of convergence in non-Hermitian DMRG.
Looking beyond the numerically accessible regime, the observed broadening of the confidence intervals and the near-degeneracy of the leading eigenvalues suggest a possible crossover scenario.
In particular, we conjecture that around the Thouless time the leading eigenvalues become increasingly near-degenerate, such that at finite sizes multiple eigenvalues contribute comparably, producing the ramp, whereas in the thermodynamic limit the single largest eigenvalue eventually dominates.
This conjectured scenario is consistent with the size dependence of the Thouless time and is further illustrated by a toy-model analysis~\cite{supplementaryMaterial}.

\section{Summary and outlook}

We propose SeTN, a tensor network approach for studying chaotic dynamics in general disordered quantum systems. 
By encoding disorder into an auxiliary layer and performing the disorder averaging separately, SeTN restores spatial translational invariance and enables accurate simulations directly in the thermodynamic limit through a compact MPO transfer matrix.From the singular value spectrum, we derive and numerically confirm a universal criterion for efficient disorder encoding, $ n \gg \alpha^{2} t^{2} $, which indicates that encoding is more efficient under weak disorder, a regime that is typically chaotic and challenging for conventional TN methods, making disorder-averaged transfer matrices numerically feasible for generic Hamiltonians.

We compute the leading eigenvalue of the SeTN-derived transfer matrix for the SFF in the disordered TFIM, and show that it accurately captures the scaling behavior of the SFF within the accessible time range.
Based on the behavior observed in this regime, we conjecture that the crossover from single-eigenvalue dominance to random-matrix-theory behavior is associated with an increasing near-degeneracy among the leading transfer-matrix eigenvalues.
Clarifying how this near-degeneracy develops and how it controls the emergence and system-size dependence of the Thouless time remains an important open problem for future work.

SeTN is broadly applicable to observables with multilayer TN representations, including Rényi entropies~\cite{chanSolutionMinimalModel2018c,zhouEmergentStatisticalMechanics2019}, out-of-time-ordered correlators (OTOCs)~\cite{nahumOperatorSpreadingRandom2018c,vonkeyserlingkOperatorHydrodynamicsOTOCs2018a,chanSolutionMinimalModel2018c}, and spin-glass order parameters~\cite{huseLocalizationprotectedQuantumOrder2013a,pekkerHilbertGlassTransitionNew2014a}, offering a powerful platform for investigating the impact of disorder on dynamical phenomena such as diffusion and hydrodynamics~\cite{khemaniOperatorSpreadingEmergence2018c,rakovszkyDiffusiveHydrodynamicsOutofTimeOrdered2018c}.
SeTN also serves as a bridge between local averaging schemes and other concepts, such as Feynman-trajectory picture~\cite{vonkeyserlingkOperatorHydrodynamicsOTOCs2018a,zhouEmergentStatisticalMechanics2019,zhouEntanglementMembraneChaotic2020}, influence functionals~\cite{nguyenCorrelationFunctionsTensor2024,sonnerInfluenceFunctionalManybody2021a,yeConstructingTensorNetwork2021,leroseInfluenceMatrixApproach2021b} and entanglement barriers~\cite{rathEntanglementBarrierIts2023}.
Overall, our construction provides a unified framework for exploring the interplay of locality, unitarity, disorder, and chaos in static many-body systems. 
In this sense, SeTN complements recent developments such as emergent symmetries in disorder-averaged dynamics~\cite{erpeldingExploitingEmergentSymmetries2025a} and translationally invariant MPO descriptions of thermal ensembles~\cite{vervoortExtractingAverageProperties2025}, and opens multiple avenues for future investigation.

\textit{Acknowledgment.}--- This work was supported in part by National Natural Science
Foundation of China Grants No. 12474353 and Beijing Natural Science Foundation (Grant No. 1232025). 
SJR acknowledges supports from Peng Huanwu Visiting Professor Program, Chinese Academy of Sciences, and Academy for Multidisciplinary Studies, Capital Normal University.
D.W. acknowledges support from the National Research Foundation, Singapore and the Agency for Science, Technology and Research (A*STAR) under the Quantum Engineering Programme (NRF2021-QEP2-02-P03) and from CQT Core Funding Grant CQT\_SUTD\_2025\_01.

\textit{Data availability.}--- All data are available from the authors upon reasonable request.

\bibliography{main}

\end{document}


\title{\bf \large Supplemental Material: \texorpdfstring{ \\ }{} Statistics-encoded tensor network approach in disordered quantum many-body spin systems}
\author{Hao Zhu}
\affiliation{School of Physics, Beihang University,100191,Beijing, China}
\author{Ding-Zu Wang}
\email[Corresponding author: ]{dingzu\_wang@sutd.edu.sg}
\affiliation{Science, Mathematics and Technology Cluster, Singapore University of Technology and Design, 8 Somapah Road, 487372 Singapore}
\affiliation{Centre for Quantum Technologies, National University of Singapore 117543, Singapore}
\author{Shi-Ju Ran}
\email[Corresponding author: ]{sjran@cnu.edu.cn}
\affiliation{Center for Quantum Physics and Intelligent Sciences, Department of Physics, Capital Normal University, Beijing 10048, China}
\author{Guo-Feng Zhang}
\email[Corresponding author: ]{gf1978zhang@buaa.edu.cn}
\affiliation{School of Physics, Beihang University,100191,Beijing, China}

\maketitle
\tableofcontents
\renewcommand{\thefigure}{S\arabic{figure}}
\setcounter{figure}{0}
\renewcommand{\thetable}{S\arabic{table}}
\setcounter{table}{0}
\renewcommand{\theequation}{S\arabic{equation}}
\setcounter{equation}{0}
\renewcommand{\thesection}{\Roman{section}}
\setcounter{section}{1}

\section*{\thesection\quad Generalization of the SeTN Framework}
\stepcounter{section}

In the main text, we focus on a simplified case where the clean Hamiltonian consists of nearest-neighbor interactions, i.e., $H = \sum_i H_{i,i + 1}$ and $H_{i} = \sigma^{z}_{i}$, with $\sigma^{z}_{i}$ being the Pauli-$z$ operator at site $i$.
Here, we outline how the statistics-encoded tensor network (SeTN) framework can be generalized to broader classes of Hamiltonians.

\begin{figure}[th]
    \centering
    \includegraphics[width=0.8\linewidth]{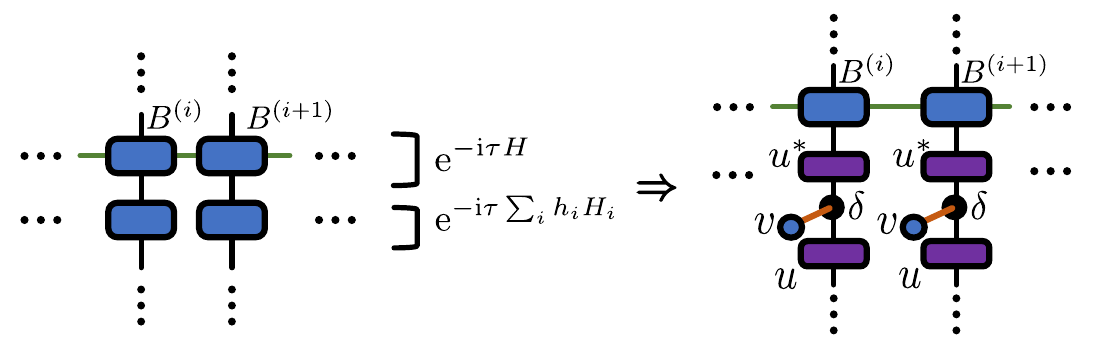}
    \caption{
        Tensor network representation of the generalized SeTN framework.
        (Left) Clean Hamiltonian evolution represented as an MPO with local tensors $B^{(n)}$.
        (Right) Representation of a general local disorder term $e^{-\mathrm{i}\tau h_i H_i}$ via eigen-decomposition.
        The combination of these elements modifies the local tensor $W$ in the SeTN construction while preserving its overall structure.
    }
    \label{fig:tn_repr_general}
\end{figure}

When $H_\text{clean}$ contains interactions beyond nearest neighbors, the corresponding step evolution operator can still be represented by matrix product operators (MPOs).
This is enabled by automata-based constructions~\cite{crosswhiteFiniteAutomataCaching2008} or methods based on the time-dependent variational principle (TDVP)~\cite{haegemanUnifyingTimeEvolution2016a}.
Specifically, we first express disorder-free term $H$ in MPO form.
By acting the evolution operator on the identity matrix (or equivalently, the infinite-temperature state), i.e., $\mathrm{e}^{ - \mathrm{i} \tau H} \cdot I$, we obtain an MPO representation of the evolution operator.
This structure is illustrated in the first row of Fig.~\ref{fig:tn_repr_general}, where we denote the MPO tensor at site $n$ by $B^{(n)}$.

For general local operators $H_i$ that are not diagonal in the computational basis, we can still apply the SeTN framework by performing an eigen-decomposition:
\begin{align*}
    (\mathrm{e}^{ -\mathrm{i}\tau h_i H_i} )_{jp}
    = u_{jq} v_{q} \delta_{qr} u_{rp}^{*}
    = v_{l} \big( u_{jq} u_{rp}^{*} \delta_{lqr} \big),
\end{align*}
as shown on the right side of Fig.~\ref{fig:tn_repr_general}.
Here, we omit the layer index $n$ and site label $(i)$ for clarity, and summation over repeated indices is implied.

With these preparations, the local tensor $W$ used in the main text is modified as follows:
\begin{align*}
    W_{j_{n}^{[i]}j_{n + 1}^{[i]}, k_{n}^{[i - 1]}k_{n}^{[i]}, l_{n}^{[i]}} =
    u_{j_{n}^{[i]}q_{n}^{[i]}} \delta_{l_{n}^{[i]}q_{n}^{[i]}r_{n}^{[i]}} u_{r_{n}^{[i]}p_{n}^{[i]}}^{*}
    B^{(i)}_{p_{n}^{[i]}j_{n + 1}^{[i]}, k_{n}^{[i - 1]}k_{n}^{[i]}}.
\end{align*}
Despite the more general setting, the tensor product structure at site $i$ remains unchanged and takes the form:
\[
    \prod_{n} v_{l_{n}^{[i]}}
    W_{j_n^{[i]}j_{n+1}^{[i]},k_n^{[i - 1]}k_n^{[i]},l_n^{[i]}},
\]
as illustrated in the third column of Fig.~1(b) in the main text.

This extension allows the SeTN framework to accommodate arbitrary local disorder terms, including those with off-diagonal components, while retaining the key structure that facilitates efficient tensor network computations.
Since this key structure originates from the underlying two-layer representation 
of disorder-averaged observables, we briefly recall this construction below.

\begin{figure}[th]
    \centering
    \includegraphics[width=0.7\linewidth]{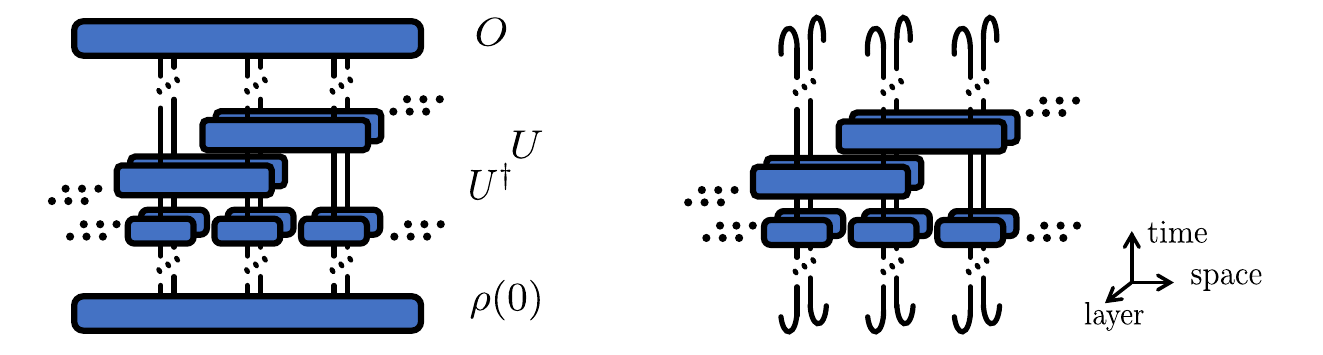}
    \caption{
        Tensor network representation of the disorder-averaged expectation values $\langle\!\langle O(t) \rangle\!\rangle$ (left) and  $K(t)$ (right).
    }
    \label{fig:tnrepr_quantities}
\end{figure}

In the main text we noted that several disorder-averaged quantities, including operator expectation values $\langle\!\langle O(t)\rangle\!\rangle = \langle \operatorname{Tr}[O\,U(t)\rho(0)U^\dagger(t)] \rangle$ and the spectral form factor 
$K(t)=\langle \operatorname{Tr}[U(t)\otimes U^\dagger(t)] \rangle$, admit a natural two-layer tensor-network representation obtained by contracting the forward and backward evolutions $U$ and $U^\dagger$.
For completeness, Fig.~\ref{fig:tnrepr_quantities} shows the corresponding tensor-network schematics of these observables, illustrating the two-layer structure that motivates the definition of the transfer matrix~$T$.
For clarity we use open boundaries, but any boundary choice leads to the same disorder-averaged structure.
For clarity we use open boundaries, although any boundary choice yields the same disorder-averaged structure.

\section*{\thesection\quad Memory efficiency in SeTN truncation}
\stepcounter{section}


In the SeTN truncation method described in the main text, one naively needs to store $n$ tensors $A^{(1)},\cdots ,A^{(n)}$, each of dimension $D^2 M^2$, where $D$ is the dimension of the internal (time) index $k$. This results in a total memory cost of $O(n D^2 M^2)$, which can be prohibitively large in practice.

This cost can be significantly reduced by exploiting the time-translation invariance of the system. After performing the MPO addition over $M$ realizations, all the tensors $A^{(n)}$ become identical for $2 \leq n \leq N-1$, except for the two boundary layers. As a result, the required memory is reduced to $O(D^2 M^2)$.

Moreover, the structure of $A^{(n)}$ allows for further simplification. For a single realization ($M = 1$), the tensor takes the form: $A^{(n)}_{l'_n l_n, 11} = v_{l'_n}^* v_{l_n}$ while for $M > 1$, corresponding to $M$ independent realizations, we have:
\begin{align*}
A^{(n)}_{l'_n l_n, j_{n-1} j_n} = v_{l'_n}^{*(j_n)} v_{l_n}^{(j_n)} \delta_{j_{n-1}, j_n},
\end{align*}
where $v_{l_n}^{(j_n)}$ denotes the vector associated with the $j_n$-th realization.

Using this structure, the contraction in the iterative truncation procedure becomes:
\begin{align*}
S^{(n-1)}_{p_{n-1}} V^{(n-1)}_{p_{n-1} j_{n-1}} A^{(n)}_{l'_n l_n, j_{n-1} j_n}
= S^{(n-1)}_{p_{n-1}} V^{(n-1)}_{p_{n-1} j_{n-1}} v_{l'_n}^{*(j_n)} v_{l_n}^{(j_n)} \delta_{j_{n-1}, j_n} 
= \sum_{j_n, p_{n-1}} S^{(n-1)}_{p_{n-1}} V^{(n-1)}_{p_{n-1} j_n} v_{l'_n}^{*(j_n)} v_{l_n}^{(j_n)}.
\end{align*}
In the last equation, we have explicitly written out the summation indices for clarity.

In this expression, $S^{(n-1)}$ is a vector of size $\chi_{n-1}$, $V^{(n-1)}$ is a matrix of size $\chi_{n-1} \times M$, and both $v^{(j_n)}$ vectors have dimension $D \times M$. The resulting tensor after summation has dimension $\chi_n \times M$, and the total memory cost of this contraction step is $O((\chi + D) M)$.

Therefore, the total memory complexity of the SeTN scheme is only $O(M)$, rather than $O(n D^2 M^2)$, which significantly improves computational efficiency.

\section*{\thesection\quad Analysis of the singular value spectrum in SeTN}
\stepcounter{section}

\subsection{Perturbative analysis}

This appendix presents a perturbative treatment of the MPO approximation depicted in Fig. ~1 (d).  
The MPO is first reshaped into a matrix product state (MPS) with physical dimension $ 4 $, upon which the approximation procedure can be formulated.
The procedure is summarized diagrammatically as follows:
\[
\includegraphics[angle=0,width= 0.5\linewidth]{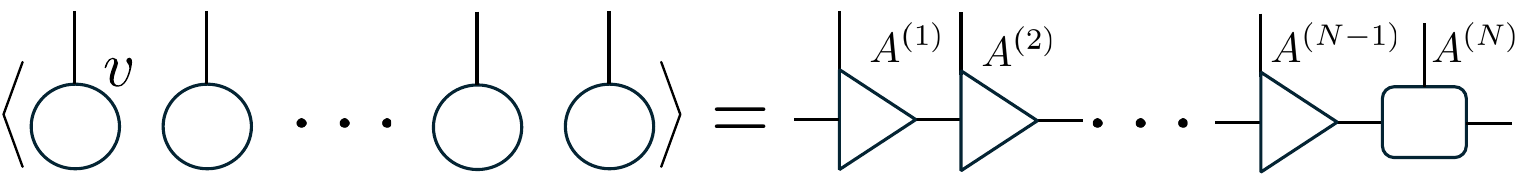}
\]
In this diagram, the circle represents a local vector \( v_k(h) \), defined by
\begin{align*}
	v_{k} =
	\begin{cases}
		1, &k = 1, 4,\\
		\mathrm{e}^{ - 2\mathrm{i} h\tau},&k = 2,\\
		\mathrm{e}^{ 2\mathrm{i} h\tau},&k = 3,
	\end{cases}
\end{align*}
while the triangles (or rectangle) correspond to local matrices \( A_k^{(n)} \), which is (not) isometric.

To be more precise, we aim as find a set of \( \{ A_k^{(n)} \} \) such that for any sequence \( \{ k_1, k_2, \dots, k_N \} \) with \( k_i \in \{1,2,3,4\} \), we have,
\begin{align*}
	&\int \mathrm{d}h \, P(h) v_{k_{1}} v_{k_2} \cdots v_{k_{N}} =
	\sum_{j_1,j_2,\cdots ,j_{N + 1} = 1}^{m} \left( A_{k_{1}}^{(1)} \right)_{j_1j_2} \left( A_{k_{2}}^{(2)} \right)_{j_2j_3}  \cdots \left( A_{k_{N}}^{(N)} \right)_{j_Nj_{N + 1}}.
\end{align*}
The random variable \( h \) follows a probability distribution function \( P(h) \), taken to be uniform on the interval \( [-\alpha, \alpha] \), that is, $P(h) = \frac{1}{2\alpha} \left( \Theta(h+\alpha) - \Theta(h-\alpha) \right)$
where \( \Theta(h) \) denotes the Heaviside step function.

To this end, we explicitly construct the local matrices \( A_k^{(n)} \) and verify that they reproduce the integral structure up to leading perturbative corrections.  
The matrices \( A_k^{(n)} \) are found to take the forms

\begin{equation}\label{eq:pert_res}
\begin{aligned}
A_{1,4}^{(n)} &= \left[
\begin{array}{ccc}
\frac{1}{2} + \delta_1 & 0 & \left( -\frac{1}{2} + \delta_2 \right)\frac{1}{\sqrt{n (2 n-1)}} \\
0 & \left(\frac{1}{2} + \frac{3}{10}\alpha ^2 \tau ^2 \right) \sqrt{\frac{n-1}{n}}  & 0 \\
\frac{2\alpha^4 \tau^4}{45} \sqrt{(n-1) (2 n-3)} & 0 & \frac{1}{2} \sqrt{\frac{(n-1) (2 n-3)}{n (2 n-1)}}
\end{array}
\right], \\
A_{2,3}^{(n)} &= \left[
\begin{array}{ccc}
\frac{1}{2}-\delta _1 & \pm \frac{1}{\sqrt{2n}} & \left( \frac{1}{2} + \delta_2 \right) \frac{1}{\sqrt{n (2 n-1)}} \\
\mp 4 \delta _1 \sqrt{\frac{n-1}{2}} &  \left(\frac{1}{2}-\frac{3}{10} \alpha^2 \tau^2\right)\sqrt{\frac{n-1}{n}} & \pm \sqrt{\frac{2 (n-1)}{n (2 n-1)}} \\
\frac{2\alpha^4 \tau^4}{15}\sqrt{(n-1) (2 n-3)} & \mp \frac{4\alpha^2\tau^2}{15} \sqrt{\frac{(n-1) (2 n-3)}{2 n}} & \frac{1}{2} \sqrt{\frac{(n-1) (2 n-3)}{n (2 n-1)}}
\end{array}
\right],
\end{aligned}
\end{equation}

where the perturbative corrections \( \delta_1 \) and \( \delta_2 \) are given by
\begin{align*}
	\delta_1 &= \frac{\alpha^2 \tau^2}{6} + \frac{(3 - 4n) \alpha^2 \tau^2}{45}, \\
	\delta_2 &= \frac{(4n - 3) \alpha^2 \tau^2}{6} - \frac{(24n^2 - 44n + 21) \alpha^{4} \tau^{4}}{45}.
\end{align*}

The derivation proceeds in three main steps:

Step 1: Approximate $ A^{(1)} $ by performing an eigenvalue decomposition of the averaged density matrix at the first `site',
\begin{align*}
	A^{(1)} = \text{eig}  \int \mathrm{d}h \, P(h) \ket{v}\hspace{-0.7mm}\bra{v},
\end{align*}
as illustrated in the following diagram:
\[
\includegraphics[angle=0,width= 0.2\linewidth]{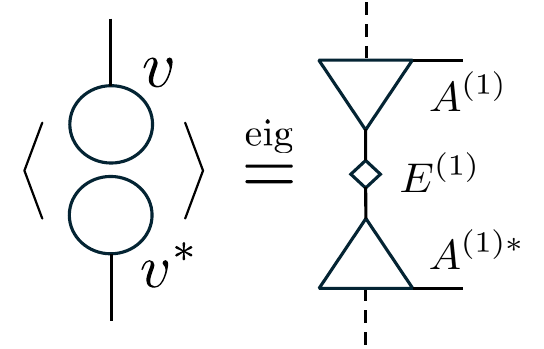}
\]
Here the dashed line denotes a bond of dimension $ 1 $, and the diamond shape represents the eigenvalues.

The integral is evaluated by expanding the vector $ v $ in powers of $ \tau $ and applying standard second-order perturbation theory.
The resulting eigenvalues are,
\begin{align*}
	& E^{(1)} = 
     \left\{4-\frac{8 \alpha ^2 \tau ^2}{3} + \frac{16\alpha ^4 \tau ^4}{9}, \frac{8 \alpha ^2 \tau ^2}{3}-\frac{32 \alpha ^4 \tau ^4}{15},\frac{16\alpha ^4 \tau ^4}{45}, 0 \right\},
\end{align*}
and the corresponding eigenvectors are given by,
\begin{align*}
	A^{(1)}_{1,4} = &
	\left[
	\begin{array}{ccc}
		\frac{1}{2} + \frac{\alpha^2 \tau^2 }{6} - \frac{\alpha^{4} \tau^{4}}{45}  & 0 & -\frac{1}{2} + \frac{\alpha^2 \tau^2 }{6} - \frac{\alpha^{4}\tau^{4}}{45} \\
	\end{array}
	\right], \\
	A^{(1)}_{2,3} = & 
	\left[
	\begin{array}{ccc}
		\frac{1}{2} - \frac{\alpha^2 \tau^2 }{6} + \frac{\alpha^{4}\tau^{4}}{45} & \pm \frac{1}{\sqrt{2}}  & \frac{1}{2} + \frac{\alpha^2 \tau^2 }{6} - \frac{\alpha^{4}\tau^{4}}{45}
	\end{array}
	\right].
\end{align*}
These expressions agree with the form shown in Eq.~\ref{eq:pert_res} at $ n = 1 $, and serve as the initial condition for the recursion.

Step 2: Approximate $ A^{(2)} $ based on the previously obtained $ A^{(1)} $. 
To construct the density matrix required for eigenvalue decomposition, we introduce the renormalized tensor $ w^{(2)} $, defined as,
\begin{align}\label{first_w}
	w^{(2)}_{i} &= \sum_{k}^{} v_{k} \left( A^{(1)}_{k} \right)_{1,i}^*.
\end{align}
The matrix $ A^{(2)} $ is then obtained by performing an eigenvalue decomposition of the averaged outer product,
\begin{align*}
	A^{(2)} &= \text{eig} \int \mathrm{d}h \, P(h) | w^{(2)} \otimes v \rangle \hspace{-0.5mm}\langle w^{(2)} \otimes v |,
\end{align*}
illustrated diagrammatically as,
\[
\includegraphics[angle=0,width= 0.25\linewidth]{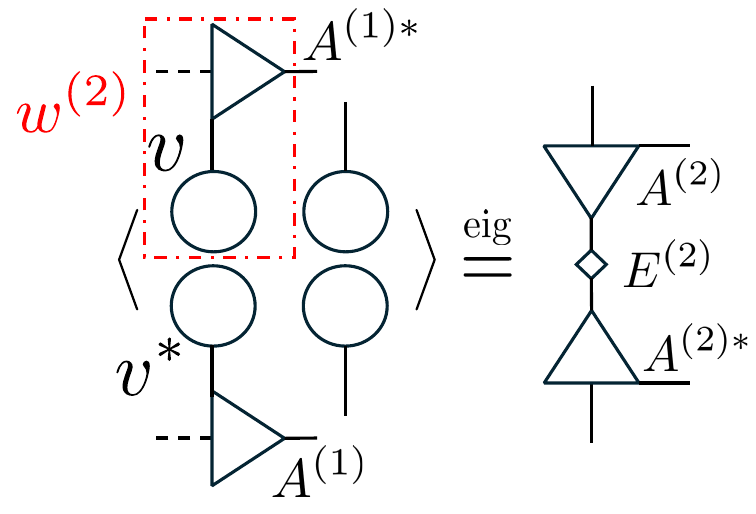}
\]

As in Step 1, the integral is evaluated perturbatively to order $ O(\tau^{4}) $.
The resulting eigenvalues are
\begin{align*}
	E^{(2)} = &\left\{16-\frac{64 \alpha ^2 \tau ^2}{3} + \frac{1616 \alpha ^4 \tau ^4}{45},
    \frac{64 \alpha ^2 \tau ^2}{3}-\frac{128 \alpha ^4 \tau ^4}{3},\frac{128 \alpha ^4 \tau ^4}{15},0 \times 9 \right\},
\end{align*}
where $ 0 \times 9 $ indicates ninefold degenerate zero eigenvalues.

The corresponding eigenvectors again match the structure given in Eq.~\ref{eq:pert_res}, now with $ n = 2 $, thereby validating the recursion result up to this order.

Step 3: We now have sufficient information to establish the general recursion relation $ (w^{(n)}, A^{(n)}) \mapsto (w^{(n + 1)}, A^{(n + 1)}) $, given by,
\begin{align}
	w^{(n + 1)}_{j} &= \sum_{k,i}^{} w_{k}^{(n)} v_{i} \left( A^{(n)}_{k} \right)_{i,j}^*,\\
	\label{eq:recursive_relationship_2}
	A^{(n + 1)} &= \text{eig} \int \mathrm{d}h \, P(h) | w^{(n + 1)} \otimes v \rangle \hspace{-0.5mm}\langle w^{(n + 1)} \otimes v |.
\end{align}
The eigenvalue decomposition is depicted diagrammatically as,
\[
\includegraphics[angle=0,width= 0.45\linewidth]{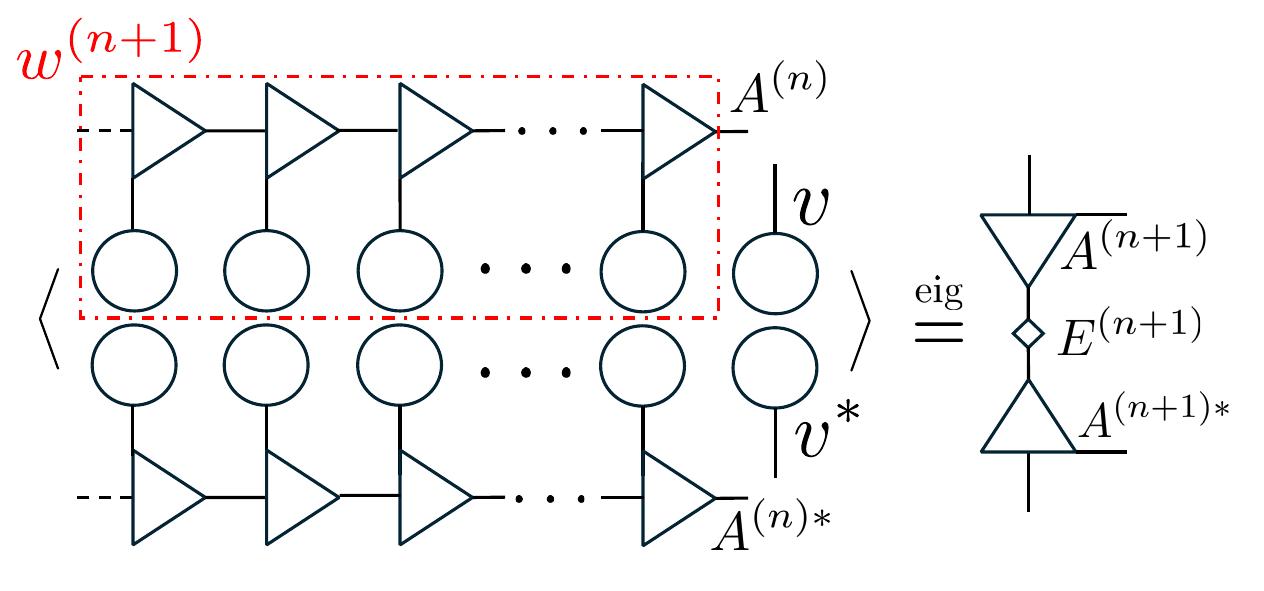}
\]

As in Step 2, the eigenvalue decomposition of the $ 12 \times 12 $ matrix is carried out perturbatively up to $ O(\tau^{4}) $.
The resulting eigenvectors allow us to propagate the recursion for $ w^{(n)} $.

We assume the general structure of $ w^{(n)} $ takes the form
\begin{align*}
	w^{(n)} = 
	\begin{bmatrix}
		w_{10}^{(n)} + w_{12}^{(n)}\tau^2 + w_{14}^{(n)}\tau^{4} \\
		\mathrm{i} w_{21}^{(n)} \tau + \mathrm{i}w_{23}^{(n)}\tau^{3} \\
		w_{32}^{(n)}\tau^2 + w_{34}^{(n)} \tau^{4} \\
	\end{bmatrix}
\end{align*}
which reflects the parity structure inherited from the expansion in $ \tau $.

From the perturbative eigenvalue decomposition, we derive recursive relations for the expanding coefficients, such as,
\begin{align*}
	w_{10}^{(n + 1)} = 2 w_{10}^{(n)},\quad w_{12}^{(n + 1)} = 2 w_{12}^{(n)} - \frac{h^2}{2} w_{11}^{(n)}.
\end{align*}
and so on.
While the full list of recursive expressions is algebraically tedious, the coefficients admit closed-form expressions:
\begin{align}
	\begin{cases}
		w_{11}^{(n)} = 2^{n}\\
		w_{12}^{(n)} = - n 2^{n} h^2 /4\\
		w_{14}^{(n)} = n 2^{n} h^2 \big(h^2 (3n - 1) + 2\alpha^2 (2n - 1)\big)/ 96\\
		w_{21}^{(n)} = - 2^{n} h \sqrt{n/2}\\
		w_{23}^{(n)} = 2^{n} h^{3} \sqrt{n/2} (3n - 1)/12\\
		w_{32}^{(n)} = 2^{n} \sqrt{n(2n - 1)} (\alpha^2 - 3h^2)/12\\
		w_{34}^{(n)} = 2^{n} \sqrt{n(2n - 1)} \big(15h^{4}(3n - 2) - 15 h^2 n \alpha^2 - 2(4n - 3)\alpha^{4}\big)/720
	\end{cases}
\end{align}
With these renormalized vectors in hand, the matrix elements in Eq. \ref{eq:pert_res} can be constructed directly from Eq.~\ref{eq:recursive_relationship_2}, completing the recursive formulation.

The key result extracted from the above procedure is the expression for the largest three eigenvalues of the reduced density matrix:

\begin{equation}\label{eq:eigval}
    \begin{aligned}
        \frac{E^{(n + 1)}}{4^{n + 1}} = \Big\{ 1 -\frac{2 (n+1)}{3} \alpha ^2 \tau ^2 + \frac{38 n^2+43 n+20}{45} \alpha ^4 \tau ^4, \frac{2 (n+1)}{3} \alpha^2 \tau^2-\frac{4(n+1) (3 n+2)}{15} \alpha ^4 \tau ^4, \frac{4(n+1) (2 n+1)}{45} \alpha ^4 \tau ^4\Big\}.	
    \end{aligned}
\end{equation}


This result can be directly validated by examining the residual norm
\begin{align*}
	\|  A^{(n + 1)\dagger} \bar{\rho} A^{(n + 1)} - E^{(n + 1)} \| \sim O(\tau^{5}) 
\end{align*}
where \( \bar{\rho} \) denotes the integrated density matrix appearing on the right-hand side of Eq.~\ref{eq:recursive_relationship_2}.

Moreover, Eq.~\ref{eq:eigval} enables an important scaling analysis in the limit \( \tau \to 0 \) and \( n \to \infty \) with the product \( t = n\tau \) held fixed. In this limit, only the leading eigenvalue survives, growing asymptotically as \( 4^{n+1} \), while the subleading eigenvalues are suppressed. The relative magnitudes of the second and third eigenvalues behave as
\begin{align*}
	\frac{E^{(n + 1)}_{2}}{E^{(n + 1)}_{1}}	\sim \frac{2}{3} \alpha^2 \tau t,\quad 
	\frac{E^{(n + 1)}_{3}}{E^{(n + 1)}_{1}}	\sim \frac{8}{45} \alpha^4 \tau^2 t^2,
\end{align*}
which is exact the results shown in the main text.

\subsection{Numerical analysis}

In the previous section, by solving the recursion relation, we showed that relative values of the three largest singular values of the MPO approximation take the form $ 2(\alpha^2 \tau t)/3 $ and $ 8(\alpha^2 \tau t)^{2}/25 $.
This suggests that the full singular value spectrum can be expressed as $c_{i}(\alpha^2 \tau t)^{i},\quad i = 1, 2, \cdots$ where $c_i$ are constants.
To verify this conjecture, we numerically investigate in this section the singular value spectrum by systematically varying the disorder strength $\alpha$, the evolution time $t$, and the Trotter step $\tau$.
We also show that, in the case of a Gaussian disorder distribution, the scaling structure identified above persists without modification.

\subsubsection{A. Uniform disorder}

\begin{figure}[tpb]
    \centering
    \includegraphics[width=\linewidth]{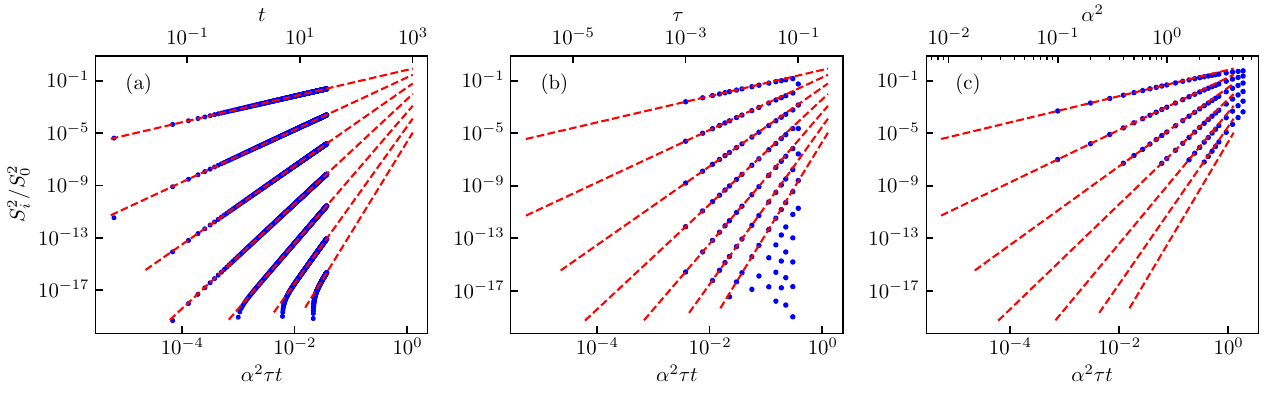}
    \caption{(Color online) 
    Uniform distribution case.
    Blue dots: The several largest squared relative singular values $(S_i(t)/S_0(t))^2$ at (a) different times $t$ with $\alpha = 0.5$ and $\tau = 0.005$, (b) different Trotter step sizes $\tau$ with $\alpha = 0.5$ at $t=15$, and (c) different disorder strengths $\alpha$ with $\tau = 0.005$ at $t=15$.
    Other parameters:$J=b=1, M=10^6$ and 
    singular values smaller than $10^{-10}$ are discarded.
    Red dashed lines: Fits to the form $c_i(\alpha^2 \tau t)^i$.
    }
    \label{fig:numerical_analysis_uniform}
\end{figure}

Figure~\ref{fig:numerical_analysis_uniform} shows the normalized squared singular spectrum at times with different parameters, with the corresponding fits to the analytic form $c_i(\alpha^2 \tau t)^{i}$ indicated by red dashed lines.
The data presented in Fig.~2(a) of the main text is selected from this set of results.

Overall, the numerical data agree well with the predicted form $c_{i}(\alpha^2 \tau t)^{i}$, supporting our conjecture.
At large values of $\alpha$ and $\tau$, deviations begin to appear, which can be attributed to the breakdown of the perturbative expansion, valid only in the regime $\alpha \tau \ll 1$.
Nevertheless, even in this regime, the numerical singular values remain smaller than the fitted values, indicating that the form $c_{i}(\alpha^2 \tau t)^{i}$ serves as an upper bound for the spectrum.

\subsubsection{B. Gaussian disorder}

\begin{figure}[tpb]
    \centering
    \includegraphics[width=\linewidth]{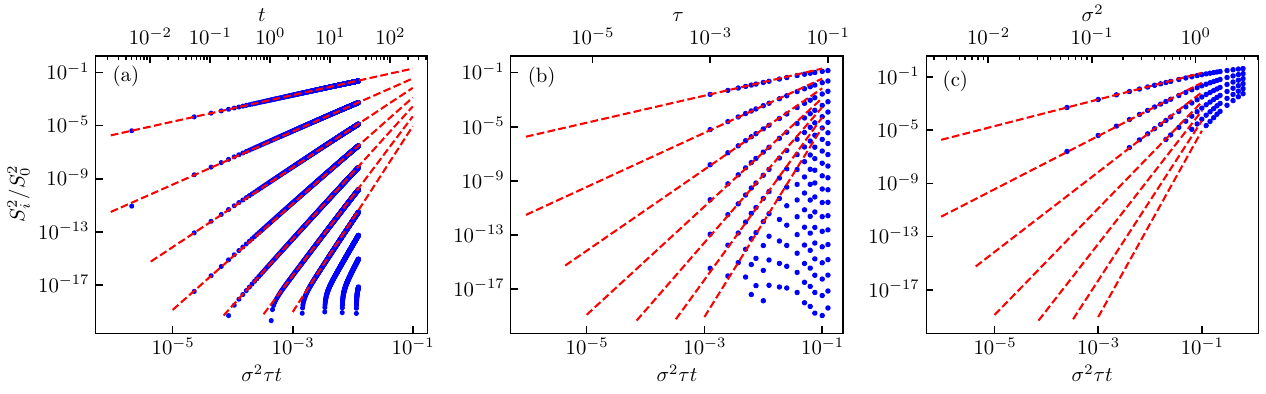}
    \caption{(Color online) 
    Gaussian distribution case.
    Blue dots: The several largest squared relative singular values $(S_i(t)/S_0(t))^2$ at (a) different times $t$ with $\sigma = 1/2\sqrt{3}$ and $\tau = 0.005$, (b) different Trotter step sizes $\tau$ with $\sigma = 1/2\sqrt{3}$ at $t=15$, and (c) different disorder strengths $\sigma$ with $\tau = 0.005$ at $t=15$.
    Other parameters:$J=b=1, M=10^6$ and 
    singular values smaller than $10^{-10}$ are discarded.
    Red dashed lines: Fits to the form $c_i(\sigma^2 \tau t)^i$.
    }
    \label{fig:numerical_analysis_gaussian}
\end{figure}

The singular-value spectra for Gaussian disorder are shown in Fig.~\ref{fig:numerical_analysis_gaussian}.
We observe that the same scaling form $ (S_i/S_0)^2 \sim c_{i}(\sigma^{2}\tau t)^{i} $ (variance $\sigma^2$) holds with excellent accuracy, confirming that the dominant dependence on disorder enters solely through the combination $ \sigma^{2}\tau t $.

Compared to the uniform case, the fitted coefficients $c_i$ are systematically larger.
This is naturally explained by the heavier tails of the Gaussian distribution:
although rare, larger disorder values contribute disproportionately to higher-order moments, and therefore enhance the weights of higher-order terms in the expansion.
Nevertheless, the singular values retain a clear monotonic decay in the index $i$, and the spectrum exhibits the same well-defined collapse when plotted against $\sigma^2 \tau t$.
Consequently, the same compressibility condition $n \gg \sigma^2 t^2 $ applies, demonstrating that the SeTN representation remains efficient for Gaussian disorder.

\subsubsection{C. Physical interpretation and numerical implications}

This combined perturbative and numerical analysis confirms that the singular values of the approximated MPO decay faster than exponentially.
As a result, SeTN with a moderate bond dimension is sufficient to represent the disorder-averaged operator $\mathrm{O}$ reliably, even at long times.

This scaling behavior also has a clear physical interpretation:
as the number of layers $n$ increases, the size of the effective Hilbert space grows exponentially, but the relevant physical information remains confined to a low-entanglement subspace that encodes the same continuous-time dynamics.
The dimensionless parameter $\alpha^2 t^2$ characterizes the total information content from disorder averaging, while the information encoded per layer scales as $\alpha^2 t^2/n=\alpha^2 \tau t$, which directly controls the singular value spectrum.

From a numerical perspective, this analysis reveals a key trade-off:
decreasing the time step $\tau$ improves the accuracy of the approximation and reduces the required bond dimension, but also increases the Trotter number $n=t/\tau$, which lengthens the MPO and raises computational cost.
Therefore, maintaining the condition $\alpha^2\tau t \ll 1$ provides a practical guideline for choosing $\tau$.
In particular, to ensure efficient SeTN compression, one should use
$n \gg \alpha^2 t^2$ or equivalently $ \tau \ll 1/(\alpha^2 t)$.

\section*{\thesection\quad Relation to Quantum-Parallel Encoding}
\stepcounter{section}
As mentioned in the main text, the SeTN framework can be viewed as a generalization of quantum-parallel encoding methods~\cite{paredesExploitingQuantumParallelism2005c,alvarezQuantumParallelismTool2008a,andraschkoPurificationManyBodyLocalization2014,hubigTimedependentStudyDisordered2019a} to continuous disorder distributions.
In this section, we elaborate on this connection.

In quantum-parallel encoding, the disorder-averaged dynamics is simulated by introducing an auxiliary quantum system that encodes discrete disorder realizations, i.e., $ P(h) = \sum_{i} p_{i} \delta(h - h^{(i)})  $, where we assume $ h $ on different sites are independent and identically distributed (i.i.d.) which is the setting considered in the main text.
The basic idea is to construct a combined Hamiltonian acting on both the physical and auxiliary systems: $ \tilde{H} = H(\hat{h}_{1},\hat{h}_{2},\cdots,\hat{h}_{L}) $, where $ \hat{h}_{i} $ are operators acting on the \textit{i}-th auxiliary system with eigenvalues corresponding to the discrete disorder values $ h^{(i)} $, i.e., $ \hat{h}_{i} \vert h^{(j)} \rangle = h^{(j)} \vert h^{(j)} \rangle $. In order to simulate the disorder-averaged dynamics, one initializes the auxiliary systems in an equal superposition state: 
\begin{align*}
    \vert \psi_{\text{aux}} \rangle = \frac{1}{N_{h}^{L/2}} 
    \Big( \sum_{i = 1}^{N_{h}}\sqrt{p_{i}}\vert h^{(i)} \rangle \Big)^{\otimes L}.
\end{align*}
By evolving the combined system $ |\Psi(0)\rangle = |\psi_{0}\rangle \otimes |\psi_{\text{aux}} \rangle $ under $ \tilde{H} $, one obtains the disorder-averaged dynamics of the physical system by $ \langle\langle O \rangle\rangle = \langle \Psi(t) | \hat{O} \otimes \hat{1} | \Psi(t) \rangle $.

\begin{figure}[tpb]
    \centering
    \includegraphics[width=0.8\linewidth]{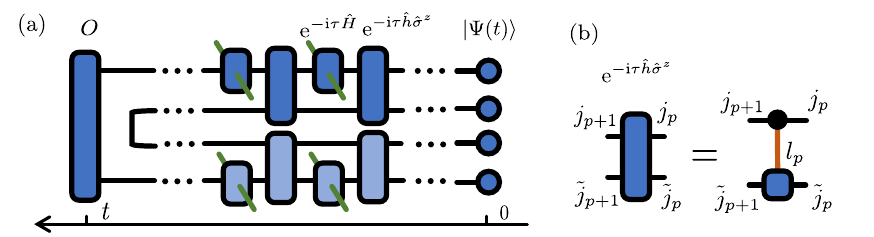}
    \caption{(Color online) 
    Trotterized representation of the disorder-averaged operator $ \langle\langle O \rangle\rangle = \langle \Psi(t) | \hat{O} \otimes \hat{1} | \Psi(t) \rangle $ in quantum-parallel encoding.
    }
    \label{fig:parallel_encode}
\end{figure}

For simplicity, we here consider the case where $ H_{i} = \sigma^{z}_{i} $ and $ P(h) = \frac{1}{2} (\delta(h + h_0) + \delta(h - h_0)) $ as in~\cite{paredesExploitingQuantumParallelism2005c} and prove that it will give the same averaged layer as in SeTN, i.e.,
\begin{align*}
    \mathrm{O}_{l'_1,...,l'_n;l_1,...,l_n} \equiv \Big\langle{\prod_{p = 1}^{n}v_{l'_p}^* v_{l_p}}\Big\rangle 
    =
    \int \mathrm{d} h \; \mathrm{e}^{ \mathrm{i} h \tau \sum_{p}^{} (l'_{p} - l_{p}) } P(h)
    =
    \cos \Big( h_0 \tau \sum_{p = 1}^{n} (l'_{p} - l_{p}) \Big).
\end{align*}
To proceed, we first decompose the $ \langle\langle O \rangle\rangle = \langle \Psi(t) | \hat{O} \otimes \hat{1} | \Psi(t) \rangle $ by Trotterization as in the main text, which is illustrated as in Fig.~\ref{fig:parallel_encode} (a).
Note that in the discrete-disorder quantum-parallel encoding, the auxiliary bond dimension is fixed at 2, which corresponds to the two allowed disorder values $\pm h_0$.

Using delta tensors, we can rewrite the layer corresponding to a single Trotter step as in Fig.~\ref{fig:parallel_encode} (b).
\begin{align*}
    (\mathrm{e}^{ - \mathrm{i} \tau \hat{h} \hat{\sigma}^{z}})_{j_{p + 1}j_{p}\tilde{j}_{p + 1}\tilde{j}_{p}} 
    = \mathrm{e}^{ - \mathrm{i} \tau \hat{h}_{\tilde{j}_{p + 1}\tilde{j}_{p}} (\hat{\sigma}^{z})_{j_{p + 1}j_{p}}} 
    = \sum_{l_p} \delta_{j_{p + 1}j_{p}l_{p}} \delta_{\tilde{j}_{p + 1}\tilde{j}_{p}} \mathrm{e}^{ - \mathrm{i} \tau h_0 \tilde{j}_{p}l_{p}}.
\end{align*}
Denoting $ A^{[p]}_{\tilde{j}_{p + 1}\tilde{j}_{p}l_{p}} = \langle \tilde{j}_{p + 1} | A^{[p]}_{l_{p}} | \tilde{j}_{p} \rangle = \delta_{\tilde{j}_{p + 1}\tilde{j}_{p}} \mathrm{e}^{ - \mathrm{i} \tau h_0 \tilde{j}_{p}l_{p}} $, the auxiliary layer can be written as
\begin{align*}
    &\quad\;\; \Big(\frac{1}{\sqrt{2}} \sum_{\tilde{j}_{1}'\in\{1,- 1\}} \langle\tilde{j}_{1}'|\Big) \prod_{p = 1}^{n} A^{[p]*}_{l_{p}'} A^{[p]}_{l_{p}} \Big(\frac{1}{\sqrt{2}} \sum_{\tilde{j}_{1}\in\{1,- 1\}} |\tilde{j}_{1}\rangle\Big) 
    = 
    \frac{1}{2}
    \sum_{\tilde{j}_{1},\tilde{j}_{1}'} \langle\tilde{j}_{1}'| 
    A_{l_1}^{[1]*} A_{l_2}^{[2]*} \cdots A_{l_2}^{[2]} A_{l_1}^{[1]} 
    |\tilde{j}_{1}\rangle 
    \\
    &= \frac{1}{2}
    \sum_{\{\tilde{j}_i\}_{i = 1}^{n + 1},\{\tilde{j}_i'\}_{i = 1}^{n}}
        \langle\tilde{j}_{1}'| A_{l_1}^{[1]*} |\tilde{j}_{2}'\rangle 
        \langle\tilde{j}_{2}'| A_{l_2}^{[2]*} |\tilde{j}_{3}'\rangle 
        \cdots 
        \langle\tilde{j}_{n}'| A_{l_n}^{[n]*} |\tilde{j}_{n + 1}\rangle 
        \langle\tilde{j}_{n + 1}| A_{l_n}^{[n]*} |\tilde{j}_{n}\rangle 
        \cdots 
        \langle\tilde{j}_{3} | A_{l_2}^{[2]} |\tilde{j}_{2}\rangle 
        \langle\tilde{j}_{2}| A_{l_1}^{[1]}  |\tilde{j}_{1}\rangle \\
    &= \frac{1}{2}
    \sum_{\{\tilde{j}_i\}_{i = 1}^{n + 1},\{\tilde{j}_i'\}_{i = 1}^{n}}
        \delta_{\tilde{j}_2'\tilde{j}_1'}\mathrm{e}^{ \mathrm{i} \tau h_0 \tilde{j}_1' l_1'}
        \delta_{\tilde{j}_3'\tilde{j}_2'}\mathrm{e}^{ \mathrm{i} \tau h_0 \tilde{j}_2' l_2'}
        \cdots 
        \delta_{\tilde{j}_{n + 1}\tilde{j}_n'}\mathrm{e}^{ \mathrm{i} \tau h_0 \tilde{j}_n' l_n'}
        \delta_{\tilde{j}_{n + 1}\tilde{j}_n}\mathrm{e}^{ -\mathrm{i} \tau h_0 \tilde{j}_n l_n}
        \cdots 
        \delta_{\tilde{j}_3\tilde{j}_2}\mathrm{e}^{ -\mathrm{i} \tau h_0 \tilde{j}_2 l_2}
        \delta_{\tilde{j}_2\tilde{j}_1}\mathrm{e}^{ -\mathrm{i} \tau h_0 \tilde{j}_1 l_1}
    \\
    &=  \frac{1}{2} 
    \sum_{\tilde{j}_1}
    \mathrm{e}^{ \mathrm{i} \tau h_0 \tilde{j}_1 \sum_p(l_p' - l_p)}
    = \cos \Big( h_0 \tau \sum_{p = 1}^{n} (l'_{p} - l_{p}) \Big).
\end{align*}
This is exactly the same as the averaged layer in SeTN obtained above.
This proof can be straightforwardly generalized to other discrete disorder distributions.

Thus, we establish that SeTN exactly recovers quantum-parallel encoding when the disorder distribution is discrete. 
In contrast, for continuous distributions quantum-parallel encoding would require an infinite-dimensional auxiliary space and becomes computationally infeasible. 
SeTN, however, remains efficient due to the exponential decay of singular values in the statistics-encoding layer, and therefore provides a natural and practically scalable extension of quantum-parallel encoding to continuous-disorder Hamiltonians.

\section*{\thesection\quad Spectral form factor of time-independent Hamiltonian}
\stepcounter{section}

In this section, we clarify the reasoning behind our choice of the spectral form factor (SFF) definition used in the main text, especially for autonomous (time-independent) systems. Although many studies do not explicitly address the subtleties of SFF definitions, a detailed discussion helps clarify the necessity of unfolding and filtering procedures, and the limitations of approximations often used.

For Floquet systems, where the dynamics is governed by a unitary Floquet operator $ F $ with eigenphases $ \{ \phi_{i} \} $, the spectral density is
\[
    \varrho(\phi) = \frac{1}{D} \sum_{i = 1}^{D} \delta(\phi - \phi_{i})
	= \frac{1}{2\pi D} \sum_{n =- \infty }^{\infty } t_{n} \mathrm{e}^{\mathrm{i}n\phi},
\]
where,
\[
	t_{n} = \sum_{i = 1}^{D} \mathrm{e}^{\mathrm{i}n\phi_{i}} = \operatorname{tr} F^{n}.
\]

The connected two-point correlation function is defined as
\[
    S(\phi,\phi') \equiv \overline{\varrho(\phi)\varrho(\phi')} - \overline{\varrho(\phi)}\;\overline{\varrho(\phi')},
\]
where the overline $ \overline{O} = \int \mathrm{d}^{D}\{ \phi_{i} \} \, P(\{ \phi_{i} \} ) O $ ensemble or disorder averaging. Assuming exchange symmetry among the eigenphases which is true in Floquet system, it follows that
\[
	\overline{t_{n}t_{n'}} = \delta_{n',- n} \overline{|t_{n}|^2},\quad\quad  \overline{\varrho(\phi)} = 1/2\pi.
\]
Then, the two-point correlation function becomes,
\[
  S(\phi,\phi') 
  = \frac{1}{2\pi^2 D^2 } \sum_{n = 1}^{\infty} \overline{|t_{n}|^2} \cos \left( n(\phi - \phi') \right).
\]
and the spectral form factor is defined as the Fourier coefficients:
$$
  K(n) = \overline{|t_{n}|^2} = \overline{\sum_{i,j = 1}^{D} \mathrm{e}^{\mathrm{i}n(\phi_{i} - \phi_{j})}}.
$$

For time-independent Hamiltonians $ H $, the eigenphases $ \{ \phi_{i} \}  $ are replaced by eigenenergies $ \{ E_{i} \}  $, which lack exchange symmetry and generally exhibit a non-uniform local spectral density. 
In practice, to extract meaningful correlations (and thus determine timescales like the Thouless time), it is common practice to unfold the spectrum, rescaling eigenenergies $ E_{i} \to \epsilon_{i} $ to achieve a uniform average spacing~\cite{gomezMisleadingSignaturesQuantum2002,haakeQuantumSignaturesChaos2018,gharibyanOnsetRandomMatrix2018a,suntajsQuantumChaosChallenges2020b,sierantPolynomiallyFilteredExact2020}.
Additionally, edge effects due to finite bandwidth are mitigated by applying a smooth filter function $ \rho(\epsilon) $ (e.g., a Gaussian):
\begin{equation}
    K(t) = \overline{ \left| \sum_{i = 1}^{D} \rho(\epsilon_{i})\mathrm{e}^{ - 2\pi\mathrm{i}t\epsilon_{i}}\right|^2 },
\end{equation}
where $ \{ \epsilon_{1} \leq \epsilon_{2} \leq \cdots \leq \epsilon_{D} \} $ are the unfolded eigenvalues.

This definition of the SFF is essential for accurately determining the Thouless time in autonomous systems, as it accounts for the non-uniform local density of states and mitigates edge effects. 
However, as argued in~\cite{gharibyanOnsetRandomMatrix2018a}, sample-to-sample fluctuations in the eigenvalue density do not qualitatively alter the ramp structure of the SFF. 
The primary impact is a rounding of the sharp ramp-to-plateau transition, as different spectral regions saturate at different times. 
The magnitude of this effect depends on the spectral density $ \varrho(E) $ and the specific system studied.

This robustness of the ramp implies that density fluctuations across samples do not significantly affect the overall ramp structure, making the SFF a reliable diagnostic of the Thouless time even without unfolding. 
For simplicity, we adopt the SFF definition given in the main text and defer a detailed analysis of unfolding and filtering procedures to future work.

\section*{\thesection\quad Chaotic signature in disordered transverse-field Ising model}
\stepcounter{section}

\subsection{Mean Level Spacing Ratio}

An important feature of chaotic systems is the presence of spectral correlations, which can be quantified by the level spacing ratio $ \langle r \rangle  $~\cite{atasDistributionRatioConsecutive2013a}:
\begin{equation}
    r = \min \left\{ r_{n}, \frac{1}{r_{n}} \right\}, \quad 
    r_{n} = \frac{E_{n + 1} - E_{n}}{E_{n} - E_{n - 1}},
\end{equation}
where $ \{ E_{n} \} $ denotes the ordered eigenvalues of $ H_\text{dTFIM} $.
The average value $ \langle r \rangle $, computed over all eigenenergies, takes a characteristic value $ r_\text{GOE} \simeq 0.5307 $ for systems with time-reversal symmetry exhibiting Gaussian orthogonal ensemble (GOE) statistics, and $  r_\text{PS} \simeq 0.3863 $ for systems following Poisson statistics, typical of integrable or localized behavior.

The level spacing ratio $ \langle r \rangle $ is shown in the main panel of Fig.~\ref{fig:sff_ed} (a) as a function of disorder strength $ \alpha $ for system size $ L = 11,13 $ and $ 15 $.
It remains close to the GOE value \( r_\text{GOE} \) in the range $ 0.2 \sim 1.3 $, indicating chaotic behavior, and approaches the Poisson value $ r_\text{PS} $ for strong disorder, $ \alpha \gtrsim 2.5 $, signaling localization.
Here and in main text, we focus on the representative case of $ \alpha = 0.5 $, marked by the red star in the main panel.

For small disorder $\alpha < 0.1$ as shown in the inset of Fig.~\ref{fig:sff_ed} (a), the level spacing ratio deviates from both limiting values. 
This can be attributed to approaching the clean transverse field Ising model, where hidden symmetries need to be resolved before applying level statistics reliably.
We also observe that the chaotic region slightly broadens
with increasing system size, consistent with previous findings in other time-independent Hamiltonian systems~\cite{morningstarAvalanchesManybodyResonances2022a}. 
This trend suggests that quantum chaos remains robust in the central part of the disorder range.

\subsection{Thouless Time Scaling}

Within RMT, the GOE SFF exhibits a universal ramp-plateau structure~\cite{liuSpectralFormFactors2018b}:
\begin{equation}
    K_\text{GOE} (t) =
    \begin{cases}
      2t - t \log (1 + 2t/D), & t \leq D\\
      2D - t  \log \frac{2t/D + 1}{2t/D - 1}, & t > D
    \end{cases},
\end{equation}
where $D$ is the Hilbert space dimension, and $t_\text{H} = D$ denotes the Heisenberg time.

The ramp-plateau structure signals spectral correlations indicative of quantum chaos.
The ramp's onset time $ t_\text{Th} $, scaling with system size, reflects an intrinsic dynamical timescale characterized by Lyapunov exponents~\cite{bertiniExactSpectralForm2018,bertiniRandomMatrixSpectral2021a,chanSpectralLyapunovExponents2021}.

For systems without conserved quantities, the growth of $t_\text{Th}$ arises from domain walls in emergent statistical mechanical models, which separate expanding RMT-like regions~\cite{chanSpectralStatisticsSpatially2018c,garrattLocalPairingFeynman2021c,garrattManyBodyDelocalizationSymmetry2021a}.
In contrast, for systems with conserved quantities or local constraints, the scaling behavior of $t_\text{Th}$ is governed by the diffusive or subdiffusive transport of the conserved charges~\cite{friedmanSpectralStatisticsManyBody2019a,moudgalyaSpectralStatisticsConstrained2021a,dagManybodyQuantumChaos2023a}.

\begin{figure}[H]
	\centering
	\includegraphics[width=0.9\linewidth]{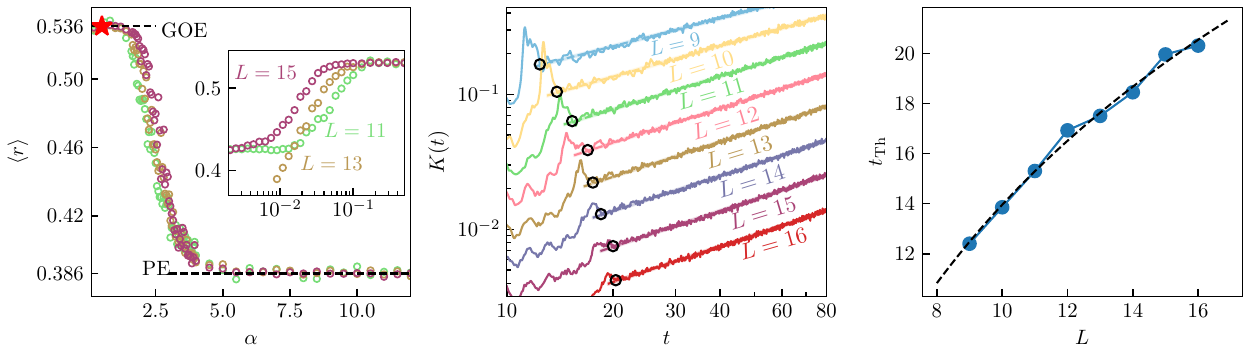}
	\caption{(Color online) Spectral form factor (SFF) of the disordered transverse field Ising model for various system sizes, zoomed into the region near the Thouless time.  Each curve is averaged over 2440 disorder realizations of $\boldsymbol{h}$, uniformly distributed with disorder strength $\alpha = 0.5$ and coupling constants $J = b = 1$. The shaded region represents a linear fit of the data between $t=20$ and $t=90$, with a 95\% confidence interval. The inset (b) shows the extracted Thouless time (indicated by the black circle in the main panel) as a function of system size.
}
	\label{fig:sff_ed}
\end{figure}

The zoomed-in SFF of dTFIM with different lengths is plotted in Fig. \ref{fig:sff_ed} (b), where we maximally calculate 2440 averages of realizations in the systems with Hilbert space dimension $ 2^{16} = 65536 $.

We fit the SFF in the region between $ 20 $ and $ 90 $ with a linear function. 
$ 95 \% $ confidence interval of the fit is shown as the shaded region.
We use the first section with the upper bound of the shaded region after the peak as an estimation of the Thouless time.
Even though it is not the most accurate way to determine the Thouless time, it is sufficient to show the log scale increase as the system size, with is consistent with many other models without conserved quantities.
The behavior of $ K(t) $ at intermediate times $ 1 \ll t \lesssim t_{\text{Th}} $ also remains an important open problem in quantum chaos.

\subsection{Effects of Boundary Conditions}

To investigate the role of boundary conditions in the scaling analysis of the spectral form factor (SFF), we repeated the calculations of Fig.~3 of the main text using periodic boundary conditions (PBC) instead of the open boundary conditions (OBC) adopted in the main text. 
Apart from the boundary condition, all parameters and numerical procedures are kept identical.

\begin{figure}[h]
    \centering
    \includegraphics[width=0.65\linewidth]{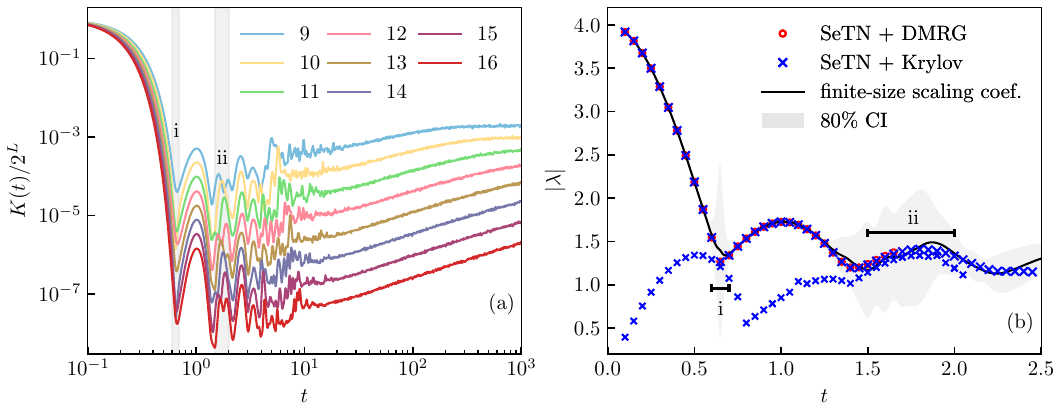}
    \caption{
        Comparison between open (OBC) and periodic (PBC) boundary conditions at disorder strength $\alpha=0.5$. 
        (a) Scaled SFF $K(t)/2^L$ for $L=9$--$16$, each averaged over 756 disorder realizations.
        The relative errors are below $20\%$ at all times and are therefore not visible on the logarithmic scale.
        The shaded bands labeled (i) and (ii) mark the first and second rebound windows. 
        (b) Magnitudes of the leading and subleading eigenvalues of the SeTN transfer matrix.  
        Horizontal bars indicate the same time windows as in panel~(a). 
        All parameters other than the boundary condition are identical to those used in the main text.
    }
    \label{fig:pbc_comparison}
\end{figure}

Figure~\ref{fig:pbc_comparison} shows the scaled SFF under PBC and the corresponding leading transfer-matrix eigenvalues, presented in the same way as in the main text. 
The two characteristic time windows associated with the first and second rebounds in the SFF, labelled (i) and (ii), are highlighted. 
Comparison with the OBC data in the main text leads to two main observations.

First, the key correspondence emphasized in Fig.~3 of the main text remains fully valid under PBC: the rebound intervals of the scaled SFF align with the time windows in which the leading and subleading transfer-matrix eigenvalues approach near-degeneracy.
Thus, the physical mechanism connecting SFF dynamics to the transfer-matrix spectrum is robust with respect to the boundary condition.

Second, when comparing the finite-size behavior across system sizes, the OBC data exhibit visibly smoother scaling over the intermediate-time regions, including and extending beyond windows (i) and (ii), whereas the PBC curves show more pronounced size dependence. 
This difference is expected to diminish as $L$ increases, and both boundary conditions yield the same characteristic timescales in the time windows relevant for the analysis in the main text. 
A detailed understanding of why the two boundary conditions behave differently would require a more systematic investigation that lies beyond the scope of the present work and is an interesting direction for future study.
But importantly, both boundary conditions exhibit the same correspondence between the SFF rebounds and the near-degeneracy of the dominant transfer-matrix eigenvalues, ensuring that the conclusions of the main text remain unchanged.

\section*{\thesection\quad Implementation of Transfer Matrix Methods}
\stepcounter{section}

To compute the leading eigenvalue of the SeTN-derived transfer matrix, we employ both the Arnoldi and non-Hermitian DMRG methods. 
In this section, we outline the technical details of our implementation, with an emphasis on exploiting the specific tensor network structure of the disordered transverse-field Ising model (disordered TFIM) after Trotter decomposition.

\subsection{Efficient Application of the Transfer Matrix}

\begin{figure}[tpb]
\centering
\includegraphics[width=0.8\linewidth]{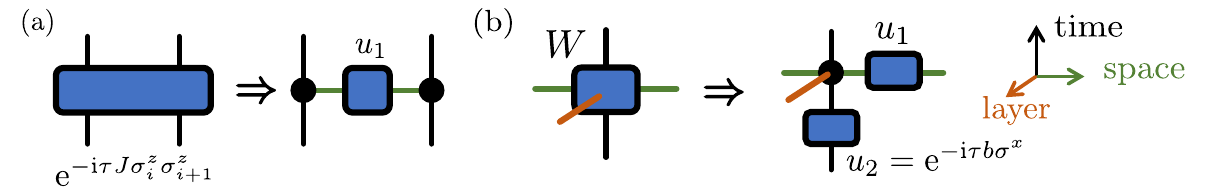}
\caption{(Color online) (a) Decomposition of the $zz$ interaction gate into Kronecker deltas and a time-direction gate $u_1$. (b) Simplified form of the $W$ tensor as a combination of local gates and delta tensors.}
\label{fig:tn_repr_sm5}
\end{figure}

The evolution operator of the dTFIM decomposes into a structure identical to that of the kicked Ising model, allowing for efficient tensor contractions. 
The clean part of the Hamiltonian, $H_{\text{clean}}$, comprises an $x$-field term and a nearest-neighbor $zz$ interaction term. 
The former yields on-site gates, while the latter is a diagonal two-site gate that can be simplified to an effective on-site gate in the time direction:
\begin{align*}
(\mathrm{e}^{ - \mathrm{i}\tau J \sigma^{z}_{i} \sigma^{z}_{i + 1}} )_{j_n^{[i]}j_{n+1}^{[i]},j_n^{[i + 1]}j_{n+1}^{[i + 1]}} &=
\mathrm{e}^{ - \mathrm{i}\tau J \sigma_{j^{[i]}_{n}} \sigma_{j^{[i + 1]}_{n}}} \delta_{j_n^{[i]}j_{n+1}^{[i]}} \delta_{j_n^{[i + 1]}j_{n+1}^{[i + 1]}} \\
&= \sum_{k_{n}^{[i]},k_{n}^{[i + 1]}}
\delta_{j_n^{[i]}j_{n+1}^{[i]}k^{[i]}_{n}}
\mathrm{e}^{ - \mathrm{i}\tau J \sigma_{k^{[i]}_{n}}  \sigma_{k^{[i + 1]}_{n}}}
\delta_{j_n^{[i + 1]}j_{n+1}^{[i + 1]}k^{[i + 1]}_{n}},
\end{align*}
where $\sigma_{j^{[i]}_{n}} = 1$ if $j^{[i]}_{n} = 1$ and $-1$ if $j^{[i]}_{n} = 2$. This structure is illustrated in Fig.~\ref{fig:tn_repr_sm5}(a), with $(u_1)_{k^{[i]}_{n}k^{[i + 1]}_{n}} = \mathrm{e}^{ - \mathrm{i}\tau J \sigma_{k^{[i]}_{n}}  \sigma_{k^{[i + 1]}_{n}}}$. The resulting $W$ tensor can thus be factorized into a five-index Kronecker delta, a time-direction gate $u_1$, and a space-direction gate $u_2$, as shown in Fig.~\ref{fig:tn_repr_sm5}(b).

\begin{figure}[tpb]
\centering
\includegraphics[width=0.9\linewidth]{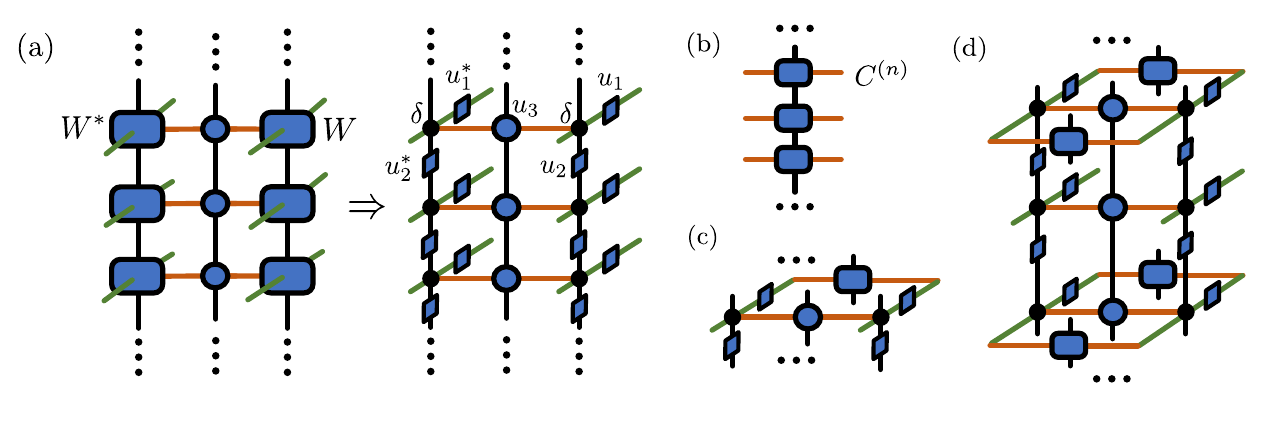}
\caption{(Color online) (a) Structure of the transfer matrix contraction. (b) MPO representation of the variational state in DMRG. (c) Efficient contraction procedure for transfer matrix application. (d) Local effective Hamiltonian for the DMRG update.}
\label{fig:tn_repr_sm4}
\end{figure}

Using this structure, we represent the transfer matrix as in Fig.~\ref{fig:tn_repr_sm4}(a), enabling efficient application to a vector without explicitly forming the full transfer matrix. 
The matrix-vector multiplication proceeds in stages: first applying $u_1$ and $u_1^*$, followed by alternating contractions with the disorder-encoding tensor $u_3$ and the evolution tensors $u_2, u_2^*$. 
This process is visualized in Fig.~\ref{fig:tn_repr_sm4}(c). 

\subsection{Krylov-Schur Implementation}

The locality of the gates allows for an efficient implementation of the Krylov-Schur algorithm~\cite{stewartKrylovSchurAlgorithmLarge2002} for extracting the dominant eigenvalues of the SeTN transfer matrix.

We initialize the Krylov iteration with an MPO of bond dimension~1 whose local tensors are identity matrices.  
Each application of the transfer matrix is carried out using the contraction procedure described above, without explicitly forming the full matrix.  
After each local contraction, we compress the resulting MPO via QR--SVD and discard singular values with relative magnitude below $10^{-14}$; the maximum allowed bond dimension is set to~$1000$.  
Vector additions appearing in the Krylov recursion are similarly implemented via MPO addition followed by compression, where singular values with relative magnitude below $10^{-12}$ are discarded.

For times $t \lesssim 2.5$, this procedure yields stable convergence of both the MPO and the extracted eigenvalues.  
For larger times, the required bond dimension grows rapidly and the Krylov iterations no longer converge within the preset tolerance.  
We expect that this limitation can be pushed to later times by further optimizing the Krylov-Schur parameters (such as the subspace dimension, truncation thresholds, and restart strategy), but we leave such systematic improvements to future work.

\subsection{Non-Hermitian DMRG with Eigenvalue Tracking}

For the non-Hermitian DMRG approach, we similarly exploit the MPO structure of the transfer matrix. 
Representing the variational state as an MPO [Fig.~\ref{fig:tn_repr_sm4}(b)], the local effective Hamiltonian takes the form shown in Fig.~\ref{fig:tn_repr_sm4}(d).

We initialize the algorithm with a random MPO of bond dimension~2 and use a relative convergence tolerance of $10^{-3}$.  
The maximum bond dimension is increased sequentially to 10, 20, 100, 200, and 1000, with each increment applied after 5 sweeps; at every local update, singular values with relative magnitude below $10^{-10}$ are discarded.  
Since the transfer matrix is non-Hermitian (as $\tau\ll1$ and the model is away from the self-dual point), we employ the Arnoldi method rather than Lanczos for diagonalizing the local effective Hamiltonian.

At intermediate and longer times, several eigenvalues acquire similar magnitudes but different complex phases, which can disrupt convergence if the target eigenvalue switches between sweeps.  
To mitigate this issue, we implement an eigenvalue-tracking scheme: after each update, we select the eigenvalue closest (in Euclidean distance) to the previously chosen one, provided its magnitude is larger.  
This continuity significantly improves stability.

With this optimization, we reliably obtain the dominant eigenvalue up to $t = 1.8$ (corresponding to $n = 40$ Trotter steps), as shown in Fig.~3(b) of the main text.  
For larger times, the required bond dimension grows rapidly and the current implementation no longer converges within the preset tolerance.  
We expect that this accessible time window can be extended by optimizing the update strategy and incorporating recent advances in non-Hermitian DMRG techniques~\cite{zhongDensityMatrixRenormalization2025a}, but a systematic investigation is left for future work.

\section*{\thesection\quad A toy model for Thouless time scaling and RMT ramp}
\stepcounter{section}

As discussed in the main text, we conjecture that around the Thouless time, the leading eigenvalues of the effective transfer matrix gradually become nearly degenerate over time, generating the linear ramp in the spectral form factor (SFF). 
At fixed time, the true largest eigenvalue eventually dominates as the system size increases.
To illustrate this idea, we introduce a simple toy model that captures the essential features of both the emergence of the RMT ramp and the growth of the effective Thouless time.

We consider a set of eigenvalues $\{\lambda_i(t)\}$ of an effective transfer matrix at time $t$,
\begin{equation}
    \lambda_i(t) = 
    \begin{cases}
        1,  & \text{for}\; i =0\\
        1 - 0.99^{\,t-i}, & \text{for}\; 0 < i \leq t\\
        0,  &\text{for}\; i > t
    \end{cases}
\end{equation}
where the subleading eigenvalues gradually approach the leading one, mimicking the flattening of the spectrum around the Thouless time, while eigenvalues with $i>t$ are set to zero to limit the number of contributing modes.
From these eigenvalues, we define the corresponding SFF for system size $L$ as
\begin{equation}
    K(t) = \sum_{i} \lambda_i(t)^L.
\end{equation}

\begin{figure}[t]
    \centering
    \includegraphics[width=0.65\textwidth]{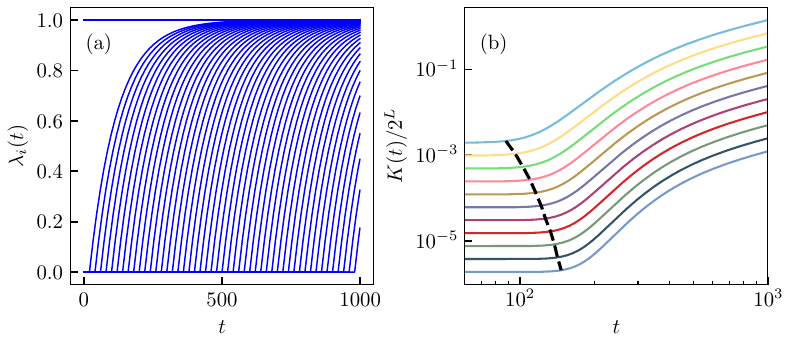}
    \caption{Toy model illustrating the conjectured transition. 
    (a) Eigenvalues $\lambda_i(t)$ as a function of $t$, with $\lambda_0 = 1$ and subleading eigenvalues $\lambda_i(t) = 1-0.99^{t-i}$ for $0 < i \le t$. 
    (b) Corresponding spectral form factor $K(t) = \sum_i \lambda_i^L$ for system sizes $L=9$ to $20$ (top to bottom), showing a linear ramp in $t$ and the growth of the effective Thouless time (black dashed line), defined as the time when subleading eigenvalues start to contribute significantly. This toy model is intended to illustrate qualitative features rather than provide quantitative predictions.
    }
    \label{fig:toy_model}
\end{figure}

Figure~\ref{fig:toy_model} illustrates the behavior of this toy model. 
Panel (a) shows the spectrum $\lambda_i(t)$ as a function of $t$, highlighting the growing near-degeneracy of the leading eigenvalues. 
Panel (b) shows $K(t)$ for system sizes $L = 9$ to $20$, demonstrating two key features: 
(i) a linear ramp in $t$, reminiscent of the RMT prediction, and 
(ii) the growth of the transient time at which the single leading eigenvalue ceases to dominate (black dashed line), reminiscent of the Thouless time behavior shown in Fig.~\ref{fig:sff_ed}(b).
Despite its simplicity, this toy model captures the qualitative features of the transition from the regime dominated by a single eigenvalue to the regime where the RMT-predicted ramp emerges, illustrating our conjecture in a clear and intuitive way.

\bibliographystyle{apsrev4-2}
\bibliography{supplement}